\documentclass[12pt,preprint]{aastex}

\def\plotfiddle#1#2#3#4#5#6#7{\centering \leavevmode
    \vbox to#2{\rule{0pt}{#2}}
    \includegraphics{#1}}
\newcommand{\Mdot}{\dot M}

\newcommand{\Mdisk}{M_{\rm disk}}
\newcommand{\Msun}{M_{\odot}}

\newcommand{\Msunperyr}{M_{\odot}\,{\rm yr}^{-1}}
\newcommand{\Lsun}{L_{\odot}}

\newcommand{\percc}{\rm \,cm^{-3}}

\newcommand{\persqcm}{\rm \,cm^{-2}}

\newcommand{\gpercc}{\rm \,g\,cm^{-3}}
\newcommand{\pers}{{\rm s}^{-1}}

\newcommand{\ccps}{{\rm cm}^{3} {\rm s}^{-1}}

\newcommand{\kms}{\,{\rm km}\,{\rm s}^{-1}}

\newcommand{\ergpers}{{\rm erg\,s^{-1}}}
\newcommand{\ergperssqcm}{{\rm erg\,s^{-1}\,cm^{-2}}}
\newcommand{\AU}{{\rm AU}}

\newcommand{\Htwo}{{\rm H_{2}}}
\newcommand{\HtwoO}{{\rm H_2O}}
\newcommand{\CtwoHtwo}{{\rm C_2H_2}}
\newcommand{\HCOp}{{\rm HCO^{+}}}
\newcommand{\COtwo}{{\rm CO_2}}
\newcommand{\CHthree}{{\rm CH_3}}
\def\micron{\hbox{$\,\mu$m}}

\shorttitle{Spitzer Spectroscopy of TW Hya}
\shortauthors{Najita et al.}

\begin{document}


\title{Spitzer Spectroscopy of the Transition Object TW Hya} 


\author{Joan R. Najita}  
\affil{National Optical Astronomy Observatory, 950 N. Cherry Ave., Tucson, AZ 85719} 

\author{John S. Carr}
\affil{Naval Research Laboratory, Code 7211, Washington, DC 20375}

\author{Stephen E. Strom}
\affil{National Optical Astronomy Observatory, 950 N. Cherry Ave., Tucson, AZ 85719} 

\author{Dan M. Watson}
\affil{University of Rochester, Department of Physics and Astronomy, University of Rochester, Rochester, NY 14627} 

\author{Ilaria Pascucci}
\affil{Space Telescope Science Institute, 3700 San Martin Drive, Baltimore, MD 21218}

\author{David Hollenbach} 
\affil{SETI Institute, 515 North Whisman Road, Mountain View, CA 94043}

\author{Uma Gorti}
\affil{SETI Institute, 515 North Whisman Road, Mountain View, CA 94043; NASA Ames Research Center, Moffett Field, CA 94035}

\author{Luke Keller} 
\affil{Department of Physics, Ithaca College, Ithaca, NY 14850}



\begin{abstract}
We report sensitive {\it Spitzer} IRS spectroscopy in the 
$10-20\micron$ region of TW Hya, a nearby T Tauri star. 
The unusual spectral energy distribution of the source, 
that of a ``transition object'', indicates that the circumstellar 
disk in the system has experienced significant evolution, 
possibly as a result of planet formation.  
The spectrum we measure is strikingly different from that 
of other classical T Tauri stars reported in the literature, 
displaying no strong emission features of $\HtwoO$, 
$\CtwoHtwo,$ or HCN.  
The difference indicates that the inner planet formation region 
($\lesssim 5$\,AU) of the gaseous disk has evolved physically 
and/or chemically away from the classical T Tauri norm. 
Nevertheless, TW Hya does show a rich spectrum of emission features of   
atoms (HI, [NeII], and [NeIII]) and 
molecules ($\Htwo,$ OH, $\COtwo,$ $\HCOp,$ and possibly $\CHthree$), 
some of which are also detected in classical T Tauri spectra. 
The properties of the neon emission are consistent with an origin 
for the emission in a disk irradiated by X-rays (with a possible 
role for additional irradiation by stellar EUV). 
The OH emission we detect, which also likely originates in 
the disk, is hot, arising from energy levels up to 23,000\,K 
above ground, and may be produced by the UV photodissociation of   
water. 
The HI emission is surprisingly strong, with relative strengths 
that are consistent with case B recombination.  
While the absence of strong molecular emission 
in the 10--20$\micron$ region 
may 
indicate that the inner region of the gaseous disk has 
been partly cleared by an orbiting giant planet, 
chemical and/or excitation effects may be responsible instead. 
We discuss these issues and how our results bear on our 
understanding of the evolutionary state of the TW Hya disk. 
\end{abstract}





\section{Introduction}

Spectroscopy with the {\it Spitzer Space Telescope} has opened a new 
window on the gas in the planet formation region of circumstellar 
disks ($< 5$\,AU).  It has revealed 
the presence of organic molecules in this inner region of the disk 
(Lahuis et al.\ 2006b; Carr \& Najita 2008) and 
shown that water emission from disks is both more commonly occuring and 
extends to larger radii than previously known from ground-based 
observations (Carr \& Najita 2008; Salyk et al.\ 2008). 
{\it Spitzer} spectroscopy has also 
demonstrated that [NeII], which was predicted to arise from 
irradiated disk atmospheres (Glassgold et al.\ 2007), 
is commonly present in spectra of T Tauri stars (Pascucci 
et al.\ 2007; Lahuis et al.\ 2007; Espaillat et al.\ 2007; Guedel et al.\ 2009). 
These diagnostics offer the opportunity to study the evolution 
of gaseous disks and to obtain new insights into the processes by 
which disks evolve. 

Transition objects are an interesting class of sources to study 
in this context.  Their spectral energy distributions (SEDs) are 
interpreted as arising from an optically thin inner disk 
(within an opacity ``hole'' of radius $R_{\rm hole}$) 
that is surrounded by an optically 
thick outer disk (beyond $R_{\rm hole}$). 
Such an SED can arise in several ways, as a consequence 
of grain growth (e.g., Strom et al.\ 1989; Dullemond \& Dominik 2005), 
giant planet formation (e.g., Skrutskie et al.\ 1990; 
Marsh \& Mahoney 1992; Lubow et al.\ 1999; 
Bryden et al.\ 1999), or 
disk photoevaporation (Clarke et al.\ 2001; Alexander et al.\ 2006). 
Transition objects might arise in any of these ways, and the 
evolutionary path for any given T Tauri star, 
which may or may not involve a stint as a transition object,  
may depend on initial conditions such as the initial disk mass. 

The above processes do make different predictions for the stellar 
accretion rates and disk masses under which a transitional  
SED will arise (Najita et al.\ 2007a; Alexander \& Armitage 2007). 
As a result, stellar accretion rates and disk masses have been
employed previously in diagnosing the nature of transition objects
(Najita et al.\ 2007a; Cieza et al.\ 2008; Alexander \& Armitage
2009).

Because these processes also make different predictions for the 
structure of the gaseous disk,  
studying the gaseous disks of transition  objects is potentially 
another tool that can be used to distinguish among these 
possibilities (Najita et al.\ 2007a, 2008).
This approach, the one taken here, may therefore offer insights 
into how planets form and disks dissipate. 

At a distance of $51$\,pc (Mamajek 2005), TW Hya is the nearest 
T Tauri star and a well-known transition object. 
Its SED indicates that the region of the 
disk within $R_{\rm hole} \sim 4$\,AU of the star 
is optically thin in the continuum 
(Calvet et al.\ 2002; Uchida et al.\ 2004).  
Higher angular resolution interferometric results at 7mm 
(Hughes et al.\ 2007) 
and at $10\micron$
(Ratzka et al.\ 2007) 
further confirm a deficit of 
continuum emission from the central region of the disk. 
TW Hya is also an unusual T Tauri star in other respects.  
Despite its advanced age ($\sim 10$ Myr), 
it has a massive disk (Wilner et al.\ 2000; Calvet et al.\ 2002) 
and is still accreting 
(Muzerolle et al.\ 2000; 
Johns-Krull et al.\ 2000; 
Herczeg et al.\ 2004; 
Alencar \& Batalha 2002).  
TW Hya is also a source of energetic UV emission 
(Herczeg et al.\ 2004; Bergin et al.\ 2004)
and emits strongly in X-rays (Kastner et al.\ 1999; 
Stelzer \& Schmitt 2004; Robrade \& Schmitt 2006). 

Spectroscopy of TW Hya taken with the {\it Spitzer} Infrared 
Spectrograph (IRS) has been reported previously,  
with an emphasis on the shape of the mid-infrared continuum and 
what it says about the 
radial distribution of the dust opacity in the disk (Uchida et al.\ 2004). 
Based on a reanalysis of the same data, 
Ratzka et al.\ (2007) reported the detection of emission lines of 
HI 6--5, HI 7--6, and [NeII]. 
The [NeII] emission from TW Hya has been studied from the ground 
at high spectral resolution by Herczeg et al.\ (2007) and 
Pascucci \& Sterzik (2009). 

Here we report further IRS spectroscopy of TW Hya obtained at higher 
signal-to-noise in two later epochs.
The spectra show a rich suite of emission features, both 
atomic and molecular, some of which have been detected before  
in spectra of classical T Tauri stars.  We discuss the 
physical processes that may be responsible for the emission (section 3). 
The spectra are also strikingly different from the spectra of 
classical T Tauri stars, in that they lack strong emission 
features of $\HtwoO$, $\CtwoHtwo$, and HCN.  This suggests that  
the inner gaseous disk has evolved physically and/or chemically 
away from the classical T Tauri norm. 
We discuss how these differences bear on our understanding of 
the evolutionary state of the disk (section 4.1). 
We also discuss where in the system the detected emission 
features may arise (section 4.2).

\section{Observations and Data Reduction}

TW Hya was observed twice with the Short-High (SH) module of the 
IRS (Houck et al.\ 2004) on board
the {\it Spitzer Space Telescope} (Werner et al.\ 2004) 
as part of the program GO 30300.  
A first epoch spectrum (hereinafter SH1) was acquired in 2006.
A second, longer integration spectrum (hereinafter SH2) was acquired in 2008.
An earlier, shorter integration spectrum that pre-dates both of
the spectra from our GO program was acquired
in 2004 as part of the IRS GTO program (hereinafter GTO). 
The GTO data have been presented previously
by Uchida et al.\ (2004) and Ratzka et al.\ (2007).
Table 1 gives a log of the observations.  

All three of the data sets were reduced using the 
techniques described in Carr \& Najita (2008).
The GTO dataset lacks off-target frames and consists of just
one exposure in each of the nod positions. In this case,
the identification of bad pixels relied solely on
the inspection of the single images, using a comparison
of the two nod positions as an aid.
The spectra were calibrated by creating a spectral response
function from a suite of archival standard star spectra reduced
in the same way as the target spectra.  
The spectral response function was customized for each nod position
of each data set by 
by combining those standard star spectra that minimzed the
fringing in order 20.
Subsequent application of IRSFRINGE
(Lahuis et al.\ 2006a) found little residual fringing. 
Table 2 provides estimates of
the resulting noise per pixel and the continuum level at the
wavelengths of selected spectral features in the $10-20\micron$ range.
The noise in each spectral region was estimated from the difference of
the spectra taken in the ``A'' and ``B'' beam positions at each epoch.

\section{Results}

\subsection{Continuum and Emission Lines} 

The shape of the mid-infrared continuum is similar in all 
three spectra (Figure 1). 
The continuum in the GTO spectrum is well approximated by 1.1 times 
the flux of the continuum in SH2. 
The SH1 continuum is approximately 1.3 times the SH2 continuum, 
although it is shallower  
in slope than SH2 beyond $\sim 15\micron$. 
These differences may arise in part from pointing errors
($< 20$\% photometric uncertainity in SH for the nominal {\it Spitzer}
pointing accuracy; see {\it Spitzer} Observer's Manual, version 8.0)
or true variability in the T Tauri continuum
(e.g., Muzerolle et al.\ 2009) or a combination of these.
The [NeII] line at $12.81\micron$ and 
HI lines at $11.31\micron$ and $12.37\micron$ 
are strongly in emission at each epoch. 

Notably, in none of the three epochs does TW Hya show strong 
mid-infrared 
molecular 
emission like that seen in {\it Spitzer} spectra of classical 
T Tauri stars that have been reported in the literature 
(Carr \& Najita 2008; Salyk et al.\ 2008; see also Pascucci et al.\ 2009). 
The [NeII] line luminosity is a 
convenient reference, since 
TW Hya and AA Tau (Carr \& Najita 2008) 
have similar [NeII] line luminosities, within 
a factor of 2.  
In AA Tau, the molecular emission is stronger 
than the [NeII] emission, whereas in TW Hya, the [NeII] emission is, 
by far, one of the brightest lines in the spectrum. 

Nevertheless, TW Hya does show a rich spectrum of weaker emission
features.  To illustrate this, we show in Figure 2
the average of SH1 and SH2 after subtracting a
high order polynomial fit to the continuum.
The features that are seen correspond well in wavelength with known
atomic lines (HI, NeII, NeIII) and other molecular features
($\Htwo$, OH, $\COtwo$, $\HCOp$, and possibly $\CHthree$).
The transitions that dominate the detected OH features are
listed in Table 3.
The $\COtwo$ band has been detected both in emission
(Carr \& Najita 2008; Salyk et al.\ 2008)
and in absorption (e.g., Lahuis et al.\ 2006b) from T Tauri stars.

We identify the feature at $12.06\micron$ as the 
$\nu_2$ fundamental Q-branch of $\HCOp$ (Davies \& Rothwell 1984). 
The detection of $\HCOp$ emission is interesting, and to our
knowledge this band has not been reported previously
in any young stellar object. 
There is also a weak feature observed in the SH1 spectrum at
$16.48\micron$ that may be the $\nu_2$ Q-branch of the fundamental
band of $\CHthree$ (Bezard et al.\ 1999; Feuchtgruber et al.\ 2000).
In addition, the [NeIII] line is marginally detected in the SH2 spectrum.
Figure 3 gives an expanded view of the region around the weak features
[NeIII] and $\CHthree$ in the epoch in which they were detected
(SH2 and SH1, respectively).
These features are more apparent in the individual spectra than
in the average of SH1 and SH2 (Figure 2).

Since fitting and subtracting a high order polynomial can
affect the strength of weak features (e.g., if it incorrectly fits
the local continuum), we did not measure line strengths from the
continuum-subtracted spectrum.  We instead used the continuum-subtracted
spectrum to identify the approximate wavelengths of features to be
fit, and we then measured feature strengths from the original spectrum. 
We made a least-squares fit to each feature using a model of a Gaussian 
and a quadratic approximation to the local continuum.  A Marquardt 
routine was used to find the best fit (e.g., Bevington 1969). 
Estimated errors on the line fluxes and line widths were found by 
determining the range of values enclosed by the 68\% confidence 
interval (a ``1-$\sigma$'' estimate) that is appropriate 
for the 6 parameters in the fit (e.g., Lampton et al.\ 1976).  


Tables 4 and 5 report the fluxes, Gaussian widths, and equivalent 
widths of the features in the SH1 and SH2 spectra, respectively.
Table 6 reports the fluxes and equivalent widths 
measured from fits to the features in the GTO spectrum. 
Feature widths are not reported for the GTO spectrum because 
they are not well constrained given the lower signal-to-noise 
of that spectrum. 
The fluxes of the features measured in the GTO spectrum 
(HI 9--7, HI 7--6, [NeII], $\Htwo$, and $\HCOp$) are 
consistent within the errors with the fluxes measured in SH1 
and SH2.

\subsection{Properties and Variability of Emission Features}

For the lines that appear to be unblended
(i.e., the HI lines 12--8 at $10.5\micron$, 
9--7 at $11.31\micron$,
7--6 at $12.37\micron$, 
13--9 at $14.18\micron$, 
8--7 at $19.07\micron$, 
[NeII] at $12.81\micron$, 
and the $\Htwo$ S1 line), 
the line center positions are typically within $\pm 50\kms$ of 
their rest values.
As shown in Figure 4, the widths of these lines are, for the most part, 
consistent with 
being spectrally unresolved. 

We detect a series of OH features that are each comprised of 
multiple pure rotational transitions of OH. 
Table 3 gives the approximate wavelengths of these OH groups 
and the transitions that dominate the flux we observe. 
Most of the emission arises from transitions within the $v$=0 
ground vibrational state, 
from both the $^2\Pi_{3/2}$ and $^2\Pi_{1/2}$ rotational ladders, 
with $J$=14.5 to $J$=29.5.  
Figure 3 of Tappe et al.\ (2008) gives a graphical illustration of 
the range of upper energy levels probed by these transitions.
Some low lying $v$=1 transitions may also contribute by enhancing 
the $v$=0 emission groups when there is a wavelength 
coincidence (e.g., at $14.65\micron$ and $15.30\micron$). 
The widths of the OH features are comparable to or broader than the nominal 
spectral resolution of IRS; e.g., the features at 
$14.08\micron$, 
$14.65\micron,$ 
$15.30\micron,$
$16.02\micron$,  
and $17.78\micron$ 
are spectrally resolved (Figure 4).
The increase in the feature widths with wavelength arises because the 
spread in the wavelengths of the lines that contribute to each feature 
is progressively larger at longer wavelengths.

Figure 5 shows the ratio of the line fluxes, 
equivalent widths, and Gaussian line widths 
in SH1 compared to SH2. 
The flux of the [NeII] line is similar in the two epochs (within $\sim 10$\%), 
but its equivalent width is lower (by $\sim 30$\%) 
in SH1 than in SH2 because of the 
brighter continuum in SH1 (Figure 1).  In comparison, 
the HI lines at $11.31\micron$ and $12.37\micron$ 
are stronger in both flux (by a factor of $\sim 2$) and 
equivalent width (by $\sim 30$\%) in SH1 compared to SH2.  
The variation of the HI lines suggests true 
variability. 
Neither the $12.37\micron$ HI line nor the 
[NeII] line are spatially extended with respect to the 
continuum at the spatial resolution of {\it Spitzer}; 
hence, 
while a small pointing error 
could alter the measured line flux, 
it is unlikely to alter the line 
equivalent width.
The $12.37\micron$ line is brighter 
than the [NeII] line in SH1 and fainter than [NeII] in SH2, 
also indicating true variability.
The $\Htwo$ lines have similar fluxes in both epochs, as does 
$\HCOp$. 

Some features are apparent in one epoch but not the other. 
We find in SH2, but not SH1, 
a weak feature at the wavelength of the [NeIII]$15.55\micron$ line.  
The [NeIII] line may be more apparent in SH2 because of the 
higher signal-to-noise of the SH2 spectrum. 
We find in SH1, but not SH2, weak features at the wavelength of 
the $\CHthree$ band (at $16.48\micron$) 
and the HI 14--9 line (at $12.58\micron).$ 
Finding the latter feature in SH1 but not the higher signal-to-noise SH2 
is consistent with the brighter emission observed for the 
dominant HI lines (7--6 and 9--7) in SH1 than in SH2 (Figure 1).

\subsection{HI Lines}

The flux of the HI 7--6 line, as an average of the flux in 
SH1 ($2.3\times 10^{28}\,\ergpers$) 
and SH2 ($1.3\times 10^{28}\,\ergpers$), is in good agreement 
with the HI 7--6 flux of $1.7\times 10^{28}\,\ergpers$ 
reported by Ratzka et al.\ (2007). 

The relative fluxes of the HI lines in SH1 and SH2 
can be accounted for roughly with case B 
recombination.
Figures 6 and 7 compare the HI fluxes measured in SH1 and SH2, respectively,
with the relative fluxes for case B at 
a temperature of $T=10^4$\,K and an electron density of $n_e=10^4\percc$
(Storey \& Hummer 1995).  The case B relative fluxes are similar 
(i.e., within our measurement errors)  
over a broad range in temperature ($1,000-20,000$\,K) and 
electron density ($n_e = 10^3- 10^8\,\percc$). 
As a result, the relative fluxes we measure do not constrain the 
temperature and electron density of the emitting gas.

In comparison, 
greater variation is expected in the relative line fluxes of 
lower lying HI lines over the same range in density and temperature. 
As described by Bary et al.\ (2008), the Paschen 
(Pa$\beta$, Pa$\gamma$, Pa$\delta$, and 8 through 14) and 
Brackett (Br$\gamma$ and 10 through 16) lines of T Tauri stars 
that originate from the same range of upper energy levels 
also show line ratios consistent with case B recombination.  
They interpret their line ratios as favoring 
low temperature ($\lesssim 2000$\,K) and high electron density 
($10^{9}\,\percc < n_e \lesssim 10^{10}\,\percc$).  

The mid-infrared (MIR) lines we observe are predicted to have inverted 
populations at such high densities (Hummer \& Storey 1987). 
Transitions between high lying levels typically become inverted 
at lower densities than those between low lying levels. 
So, for example, the HI 9--7 and 8--7 transitions have a 
population inversion at densities above $n_e \sim 10^8\,\percc,$
whereas the HI 7--6 transition is inverted at higher densities 
($n_e \gtrsim 10^8 - 10^9\,\percc$)  
and the HI 13--9 transition at lower densities
($n_e \gtrsim 10^7\,\percc$). 
The fact that the relative fluxes we observe in TW Hya are consistent  
with case B recombination suggests that we are not seeing 
maser emission in this source, as the predicted amplification 
factors can differ widely between transitions.   
Similarly, MIR HI transitions were not found to be masing in 
a study of MWC349 (Smith et al. 1997), a source that shows maser 
emission in longer wavelength HI lines. 

We can estimate the number of ionizing photons that are needed to produce 
the observed HI emission under the assumption of ionization equilibrium, 
i.e., that the number of ionizations per second equals the number of 
recombinations per second. 
Since the recombination line luminosity 
$L_{ul} = n_e n_p \epsilon_{ul} V$ 
(where $V$ is the optically thin emitting volume, 
$n_e$ is the electron density, and 
$\epsilon_{ul}$ is the emissivity of the transition in 
${\rm erg\,s^{-1}\,cm^3}$), 
the total recombination rate 
$\alpha_B n_e n_p V$ 
can be expressed as 
$\alpha_B\,L_{ul}/\epsilon_{ul}.$ 
For the HI 9--7 transition at $11.31\micron$ in SH2, 
$L_{ul} = 4.5\times 10^{27}\,\ergpers$. 
From Hummer \& Storey (1987),  
$\epsilon_{ul} = 3.7\times 10^{-28}\,\ergpers \,{\rm cm^{3}},$ 
and $\alpha_B = 2.6\times 10^{-13}\,\ccps$ 
for $T_e=10^4$\,K and $n_e = 10^4\,\percc$. 
For these values, the total recombination rate is 
$3.2\times 10^{42}\,\pers.$

In comparison, the ionizing photon flux of T Tauri stars is often 
estimated as $10^{41}\,\pers$ (e.g., Hollenbach et al.\ 2000), although 
Alexander et al.\ (2005) derived ionizing photon fluxes of 
$10^{41}-10^{44}\,\pers$ for various T Tauri stars based on 
CIV $\lambda 1549$ line luminosities.  Looking at the situation 
for TW Hya itself, Herczeg (2007) estimated an ionizing photon flux 
of $5\times 10^{41}\,\pers$ from the accretion 
shock onto the star based on the observed X-ray and FUV line fluxes  
and the assumption of collisional ionization, with a 
factor of 5 uncertainty in the quoted value.
The number of ionizing photons that we infer is at the upper end 
of this range, so TW Hya may emit 
enough ionizing photons to explain the mid-infrared HI line 
fluxes that we observe. 
However, if the EUV photons are sufficient to produce the 
mid-infrared HI line fluxes, they are likely to produce more 
[NeII] or [NeIII] emission than is observed (Hollenbach \& Gorti
2009; see also below). 

We can also estimate the mass of the emitting (ionized) gas, 
under the assumption of optically thin emission and constant 
density.  The mass of ionized hydrogen is 
$m_H\,n_p\,V = (m_H/n_e)(L_{ul}/\epsilon_{ul}) 
= 1.6\times 10^{31}\gpercc/n_e$.  
This corresponds to a total number of hydrogen atoms 
of $10^{55}\,\percc/n_e.$  
If we require $n_e \lesssim 10^8 \percc$ 
to avoid inverted populations, the 
emitting mass of ionized gas is $\gtrsim 10^{23}$\,g. 
In section 4, we consider 
the region of the YSO system in which such emission might 
arise.

\subsection{Neon Lines}

The range of [NeII] equivalent widths we measure is consistent with the 
equivalent width reported for this line from ground-based high resolution 
spectroscopy. 
We measure equivalent widths of 
$63\pm 10$\AA, 
$44\pm 3$\AA, and 
$65\pm 2$\AA\ 
in the GTO, SH1 and SH2 spectra, respectively. 
In comparison, Herczeg et al.\ (2007) measured at high 
spectral resolution an equivalent width of $62\pm 11$\AA\ for the 
[NeII] line using MICHELLE on Gemini. 
TW Hya was also studied at high spectral resolution by 
Pascucci \& Sterzik (2009) using VISIR on the VLT.  They 
measured a [NeII] equivalent width of 41\AA\ and 
a flux of $3.8\times 10^{-14}\,\ergperssqcm,$ 
within 65--75\% of the flux of 
$5.9\times 10^{-14}\,\ergperssqcm,$ 
$5.0\times 10^{-14}\,\ergperssqcm$ 
and $5.6\times 10^{-14}\,\ergperssqcm$ 
measured for the GTO, SH1, and SH2 spectra, respectively.  

Time variability, pointing errors, flux calibration issues, and/or 
the narrower slit of the ground-based observations may account 
for any differences between the IRS and ground-based results.
In particular, 
the $0.4\arcsec$ slit of the VISIR observations translates into 
a physical size scale of $\pm 10$\,\AU at the 50\,pc distance of 
TW Hya.  As noted by Pascucci \& Sterzik (2009), if the [NeII] 
emission arises from a disk, the narrow 
slit may exclude a reasonable fraction of the emission, since 
Glassgold et al.\ (2007) have shown that the [NeII] emitting 
region of the disk may extend out to a radius of $\sim 20$\,AU. 
We further note here that in the calculations of Meijerink et al.\ (2008) 
for their fiducial T Tauri disk, 
$\sim 75$\% of the [NeII] emission arises from within 10\,AU. 
If there is no substantial gaseous disk within the optically 
thin region of the TW Hya disk (e.g., if the disk is dynamically cleared by an 
orbiting planet), then it would be more appropriate to look at 
the emission arising from 4--10\,AU. 
The fraction of the [NeII] emission that arises from this region of 
the disk is $\sim 65$\%, comparable to the fraction
of the IRS [NeII] flux that is recovered in the VISIR data. 

The 
MICHELLE observations of Herczeg et al.\ (2007) were also made 
with a $0.4\arcsec$ slit, but in poor seeing.  So slit losses 
may be an issue there, but the observations are likely to be 
less sensitive to a difference in the spatial extent of the 
[NeII] emission relative to the MIR continuum.  That is, the  
equivalent width of the observations could be similar to that 
measured by {\it Spitzer}, as observed. 
Near-simultaneous narrow- and wide-slit ground-based [NeII] 
observations of TW Hya, taken in good seeing, 
would be useful in constraining the spatial extent of the emission.

While [NeII] is commonly detected from T Tauri stars 
(Pascucci et al.\ 2007; Espaillat et al.\ 2007; Lahuis et al.\ 2007; 
Ratzka et al.\ 2007; Guedel et al.\ 2008, 2009; Flaccomio et al.\ 2009), 
[NeIII] is rarely detected, 
with upper limits typically an order of magnitude lower 
(e.g., Lahuis et al.\ 2007). 
In addition to our possible detection of [NeIII] from TW Hya, 
the T Tauri stars Sz 102 and WL5 also show [NeIII] emission. 
Sz 102 has neon fluxes and a [NeIII]/[NeII] ratio 
($\sim 0.06$; Lahuis et al.\ 2007) that are similar to the values 
reported here for TW Hya ([NeIII]/[NeII]$\sim 0.045$). 
In contrast to TW Hya, Sz 102 drives a jet (e.g., Coffey
et al.\ 2004, where the source is known as TH 28), so   
its neon emission may have contributions from a 
both a disk and a jet.  [NeIII] is also detected from WL5, where 
the [NeIII]/[NeII] flux ratio is much higher, $\sim 0.25$ (Flaccomio et al. 2009). 

The [NeII] and [NeIII] line fluxes that we detect from TW Hya 
are in good agreement 
with predictions for gaseous disk atmospheres that 
are irradiated by stellar X-rays.  
For an X-ray luminosity of $2\times 10^{30}\,\ergpers$, which 
is roughly the X-ray luminosity of TW Hya (Kastner et al.\ 1999; 
Robrade \& Schmitt 2006), 
the models of Glassgold et al.\ predict a 
[NeII] luminosity of $1.4 \times 10^{28}\,\ergpers$ 
and a 
[NeIII] luminosity of $1.6 \times 10^{27}\,\ergpers.$ 
In comparison, we measure [NeII] and [NeIII] luminosities of 
$1.6 \times 10^{28}\,\ergpers$ 
and 
$0.7 \times 10^{27}\,\ergpers$ 
in the SH2 spectrum. 
In the Glassgold models, the [NeIII]/[NeII] ratio is small because the 
emission arises in a region with a significant abundance of 
neutral hydrogen atoms (the ionization fraction $x_e \simeq 0.001$).  
As a result, there is efficient 
charge exchange between NeIII and HI and a consequent reduction 
in the NeIII abundance.

EUV photons reaching the disk would enhance the neon emission 
above that produced by X-ray irradiation. 
For an EUV irradiated region, which is typically fully ionized 
($x_e\sim 1$), Hollenbach \& Gorti (2009) find that a small 
[NeIII]/[NeII] ratio can be produced with a soft EUV spectrum, 
e.g., a blackbody with a temperature of 30,000\,K.  A harder 
spectrum, e.g., with constant $\nu L_\nu$, would produce 
[NeIII]/[NeII] $\sim 1,$ in conflict with the observations.  
The EUV spectrum of TW Hya has not been 
measured directly, although Herczeg (2007) has estimated a 
significant  EUV flux from the accretion column based on the 
observed FUV and soft X-ray spectrum. 
Herczeg did not comment on the spectral shape of the 
predicted EUV emission.  
The impact of the emitted EUV on the disk may be limited by 
foreground HI absorption (e.g., in a magnetosphere; 
Alexander et al.\ 2004). 
Reviewing the literature, Herczeg (2007) estimated that only 
a small fraction of the EUV (equivalent to 
$10^{39}$\,photons\,s$^{-1}$) reaches the disk.  So the EUV 
contribution to the neon emission from the disk may be limited.

\subsection{Molecules}

The OH features we detect originate from very high energy levels
(upper energy levels 6,000\,K to 23,000\,K above ground)
and have large transition probabilities
(A-values of $20-70\,\pers$).
OH emission from other young stellar disks has been reported 
previously for lower rotational lines in both the ground vibrational
state (Carr \& Najita 2008) and the fundamental ro-vibrational
band (Mandell et al.\ 2008; Salyk et al.\ 2008).
The rotational temperatures derived for the emission from 
these {\it lower} rotational levels are consistent
with thermal OH emission at 500--1000\,K.
The much {\it higher} energy rotational states observed in emission
in the SH spectrum of TW Hya are more similar to the
superthermal OH emission detected by Tappe et al.\ (2008)
in the Spitzer IRS spectrum of young stellar outflow HH 211.

One possible interpretation of the highly excited OH emission 
detected from TW Hya 
is that it arises from gas that is hot and dense enough to
collisionally populate the upper levels.
The excitation conditions for such thermal OH emission would
exceed even those for the CO fundamental emission
observed from TW Hya (upper energy levels 3,000 to 6,000\,K and
A-values $\sim 10\,\pers$), which arises from gas
at $\sim 1000$\,K in the inner 1 AU of the disk 
(Salyk et al.\ 2007).
The ro-vibrational transitions of CO are generally 
considered to be a good tracer of hot and dense molecular gas
in the circumstellar environment, and the lack of a CO
component corresponding to the hotter OH emission is curious.

The lack of $\HtwoO$ emission, along with a highly excited
OH emission spectrum, is a striking difference
between TW Hya and normal classical T Tauri stars.
However, close examination of the IRS SH spectra of classical
T Tauri stars does reveal the presence of these same OH emission
features 
(Carr \& Najita, in preparation), with rotational temperatures
$\sim 4000$\,K.  This emission was not previously recognized 
because of the complexity of the dense $\HtwoO$ emission in 
which the OH emission is embedded. 
Hence, the character of the OH emission in TW Hya and that of a
classical T Tauri star such as AA Tau may be more similar than 
would appear at first glance.

A likely possibility is that such highly excited OH emission 
arises from the photodissociation of $\HtwoO$, 
which is also the interpretation given by 
Tappe et al.\ (2008) for the OH emission from HH 211.
The photodissociation of $\HtwoO$ by UV photons in
the second absorption band of water 
($\lambda = 1200-1400$\AA)
will produce OH molecules in highly
excited rotational levels but predominantly in the ground
vibrational state (Harich et al.\ 2000).
We will present a detailed analysis of the OH spectrum
of TW Hya in a following paper.

The $\Htwo$ emission that we observe is much weaker than either 
the [NeII] or HI 7--6 emission from TW Hya.  
The line luminosities are in the range found for other accreting young stars.  
In the ground-based (IRTF/TEXES) $\Htwo$ survey of Bitner et al.\ (2008), 
the $\Htwo$ S(2) line luminosity of a low mass star such as GV Tau N 
is $\sim 3 \times 10^{-6}\Lsun,$  although upper limits on the S(2) line  
luminosity for some T Tauri stars are much lower 
(e.g., $\sim 0.4 \times 10^{-6}\Lsun$ for LkCa 15).
In comparison, 
the line luminosities of the S(1) and S(2) emission that are detected 
from TW Hya in SH2 are $0.8 \times 10^{-6}\Lsun$ and 
$0.5 \times 10^{-6}\Lsun$, respectively.   
Since the S(2) line of TW Hya is blended with an OH feature, we 
assumed an OH contribution that is similar to that of 
neighboring OH features in estimating the flux of the S(2) line.  
The observed S(1) luminosity is similar to the value of 
$0.2\times 10^{-6}\,\Lsun$ that is predicted for TW Hya by 
Gorti \& Hollenbach (2008).

Interestingly, the $\Htwo$ emission from TW Hya is stronger than 
the upper limits on the $\Htwo$ flux that Bitner et al.\ (2008) 
measured for TW Hya.  
Whereas we measure line fluxes of 
$1.3\times 10^{-14}\,\ergperssqcm$ and 
$0.6\times 10^{-14}\,\ergperssqcm$ 
for the S(1) and S(2) lines (5-$\sigma$  and 3-$\sigma$ 
detections, respectively, in SH2), Bitner et al.\ (2008) 
provide 3-$\sigma$ upper limits for the S(1) and S(2) lines 
of $0.7 \times 10^{-14}\,\ergperssqcm$
and $0.6 \times 10^{-14}\,\ergperssqcm$, respectively. 
Perhaps the narrower slit 
($0.81\arcsec$ and $0.54\arcsec$ for the S1 and S2 lines, 
respectively) 
of the TEXES measurements plays a role here as well.   
The line widths that are measured for $\Htwo$ emission from 
young stars, when it is 
detected by TEXES in more distant sources ($\sim 140$\,pc), 
are $\sim 10\kms$, which is consistent with much of the 
emission arising from the 10--50\,AU region of the disk.  
As discussed in the context of the [NeII] emission from 
TW Hya (section 3.4), much of this region of the disk is 
excluded by the narrow slit in a nearby source such as TW Hya. 
As in the case of [NeII], near-simultaneous narrow and 
wide slit ground-based observations of the $\Htwo$ emission 
from TW Hya, taken in good seeing, would be useful in 
constraining the spatial extent of the emission. 

We also detect molecules other than OH and $\Htwo$ in the TW Hya spectrum 
($\COtwo$, $\HCOp$, and possibly $\CHthree$). 
The properties of these emission features will be discussed in 
greater detail in a future publication.

\section{Discussion}

\subsection{Evolutionary State of the TW Hya Disk}

As a well-studied, nearby transition object, 
TW Hya has been a touchstone for understanding the origin 
and evolutionary state of such systems. 
Transition objects have unusual spectral energy distributions 
characterized by a deficit of infrared excess at short 
wavelengths, a trait that is interpreted as indicating 
that the disk is optically thin in the continuum within 
a given radius $R_{\rm hole}$ (Strom et al.\ 1989). 
Analyses of spectral energy distributions of transition 
objects find $R_{\rm hole}$ values of a few AU to $\sim 50$\,AU 
in some cases (Espaillat et al.\ 2007).
In comparison, the spectral energy distribution of TW Hya 
indicates that its disk is optically thin within $\sim 4$ AU 
of the star (Calvet et al.\ 2002; Uchida et al.\ 2004).  
An inner hole in the dust continuum emission is detected at 7mm 
and at $10\micron$ (Hughes et al.\ 2007; Ratzka et al.\ 2007).  

Such a dust distribution could be interpreted as a result of 
grain growth (Strom et al.\ 1989; Dullemond \& Dominik 2005), 
photoevaporation (Clarke et al.\ 2001; Alexander et al.\ 2006; 
Gorti \& Hollenbach 2009; Owen et al.\ 2010),  
or a gap opened by an orbiting giant planet 
(e.g., Marsh \& Mahoney 1992; Calvet et al.\ 2002; Rice et al.\ 2003; 
Quillen et al.\ 2004; Calvet et al.\ 2005). 
These scenarios could potentially be distinguished by examining 
properties other than the SED, since the scenarios make different 
predictions for the stellar accretion rate, disk mass, and 
radial distribution of gas in the disk (Najita et al.\ 2007a; 
Alexander \& Armitage 2007). 

The first two diagnostics (stellar accretion rates and 
disk masses) were used by Najita et al.\ (2007a) 
to probe the nature of transition objects in the Taurus star 
forming region. 
They found that Taurus transition objects have higher than 
average disk masses as well as stellar accretion rates that 
are $\sim 10$ times lower than non-transition objects.  
These properties are roughly consistent with the predictions 
of theories of giant planet formation 
(e.g., Lubow et al.\ 1999; Lubow \& D'Angelo 2006). 
The high disk mass of TW Hya ($> 0.06\Msun$; Calvet et al.\ 2002) 
and its comparatively low stellar accretion rate 
($\sim 10^{-9}\Msunperyr$; e.g., Herczeg et al.\ 2004;  
Muzerolle et al.\ 2000; Alencar \& Basri 2000) 
place it in a similar region of the $\Mdot$--$\Mdisk$ plane 
as the Taurus transition objects. 

Studies of line emission from the gaseous component of  
transition disks, like that presented here, 
offer the opportunity to complement studies of stellar accretion 
rates and disk mass, by probing 
the radial distribution of gas in the disk and therefore 
the evolutionary state of the system.   
As described by Najita et al.\ (2007a, 2008): 

\noindent (1) In the grain growth and planetesimal formation scenario, the 
inner disk is rendered optically thin in the continuum, 
but the gaseous component is unaltered and would fill the region 
within $R_{\rm hole}$.  
Emission from gas {\it everywhere} within $R_{\rm hole}$ 
would produce bright emission because of the lack of continuum 
emission from same region of the disk.   

\noindent (2) If a planet has formed with a mass sufficient to open a gap 
($\sim 1 M_J$), gas will be cleared in the vicinity of 
its orbit, but gap-crossing streams, from the outer disk to 
the planet, and from the planet to the inner disk, can allow 
continued accretion onto both the planet and the star, the latter 
via the replenishment of the inner disk within $R_{\rm inner} < R_{\rm hole}$ 
(e.g., Lubow et al.\ 1999; Kley 1999; Bryden et al.\ 1999; 
D'Angelo et al.\ 2003; Bate et al.\ 2003; Lubow \& D'Angelo 2006). 
While the {\it inner disk ($r < R_{\rm inner} < R_{\rm hole}$)} might then produce gaseous emission 
lines, the low surface filling factor of gas in the 
region of the gap would produce weak to negligible emission 
because of the small projected emitting area of the accretion 
streams.   

\noindent (3) In the case of photoevaporation, gas in the region of the 
disk photoevaporation radius (few AU; Font et al.\ 2004; Liffman 2003)
is being evaporated away faster than it can be replenished by 
viscous accretion.  As a result, the inner disk is decoupled 
from the outer disk and accretes onto the star, leaving 
behind a true ``inner hole'' in the gas distribution and 
{\it no emission} from the region within $R_{\rm hole}$. 

How do the properties of the gaseous inner disk of TW Hya 
compare with these predictions? 
UV fluorescent $\Htwo$ and 
CO fundamental emission, which are both believed to probe 
the inner $\sim 1$\,AU of T Tauri disks 
(see Najita et al.\ 2007b for a review), 
have been detected from the inner disk of TW Hya 
(Herczeg et al.\ 2002; Rettig et al.\ 2004; Salyk et al.\ 2007; 
Najita et al.\ 2007b). 
The spectroastrometric study of Pontoppidan et al.\ (2008) 
further demonstrates that the CO emission arises from a 
rotating disk close to the star ($\sim 0.1$\,AU). 
Thus, the inner disk is not completely cleared of gas, a result  
that is consistent with the ongoing stellar accretion in the 
system and inconsistent with the EUV-driven photoevaporation scenario. 
(The situation is somewhat different if photoevaporation is driven
by X-rays.  Recent work by Ercolano and collaborators finds that
photoevaporation driven by X-rays rather than EUV can create a
transition-like SED at much higher disk masses and accretion rates
than in the EUV-driven photoevaporation case.  Under these conditions,
the star may continue to accrete at a measurable rate for a longer
fraction of the system lifetime after disk clearing has begun; Owen
et al.\ 2010.)

The {\it Spitzer} spectrum allows us to probe the gaseous disk 
at radii beyond the region probed by UV fluorescent $\Htwo$ 
and CO fundamental emission, 
because the features in the SH spectrum probe 
emission from cooler gas.  In the spectrum of AA Tau, 
a typical T Tauri star that possesses an optically thick inner disk, 
the molecular features detected in SH 
($\HtwoO$, OH, $\CtwoHtwo$, HCN, $\COtwo$), 
probe gas with 
temperatures of 300-600\,K and disk emitting areas corresponding 
to a few AU in radius, i.e., 
the inner planet formation region of the disk (Carr \& Najita 2008).

One of the most striking aspects of the {\it Spitzer} spectrum of TW Hya 
is the weak molecular emission compared to that seen in 
classical T Tauri stars such as AA Tau (Carr \& Najita 2008; 
Salyk et al.\ 2008;  
Carr \& Najita, in preparation; Pontoppidan et al., in preparation). 
The scenario of a gap carved in the gaseous disk by an 
orbiting giant planet would appear to naturally lead to the  
suppression of the molecular emission from the inner disk 
region, like that observed.  With the size of the optically thin 
region in the TW Hya system ($\sim 4$\,AU) 
suggesting that an orbiting planet would reside at a few AU, 
we might expect the inner few AU region of the disk to be mostly 
cleared of gas. 

In comparison, because it suggests a continuous gaseous disk 
with no continuum opacity in the inner region, 
the grain growth scenario might predict similar or stronger 
mid-infrared 
molecular emission than would be observed in a continuous disk 
of gas and dust such as AA Tau, as long as  
the molecular abundances of the inner disk are not significantly 
altered by the optically thin nature of the inner disk. 
Photodissociation may alter the molecular abundances in the 
disk, although molecular shielding (e.g., by CO, $\Htwo$, OH, 
and $\HtwoO$) may limit its impact.
Another caveat is that molecules may be abundant in the inner 
disk of TW Hya, but molecular emission might not be excited 
to the level seen in AA Tau because of the lower accretion 
rate of TW Hya compared to typical T Tauri stars.  
A low accretion rate may imply a reduced rate of mechanical 
heating in the disk atmosphere.   A low rate of mechanical heating 
can reduce the column density of warm water in the disk atmosphere 
(Glassgold et al.\ 2009) and lead to weak water emission.

There is observational motivation to consider these issues. 
Photodissocation by stellar UV photons may play a role in 
reducing molecular abundances in the disk, and the hot 
OH we see may provide evidence for the photodissociation 
of water in this system (sections 3.5 and 4.2).  
The lack of grains in a gas-rich inner disk may also 
limit the molecular abundances in this region.  Glassgold et al.\ (2009) 
have shown that processes such as the formation of $\Htwo$ 
on warm grain surfaces and mechanical heating associated with 
accretion can significantly 
enhance the abundance of water in the warm atmospheres of 
inner disks 
up to the levels observed in classical T Tauri disks. 
While it is yet unclear 
whether either of these is the dominant or critical factor
in explaining the observed water abundances in classical 
T Tauri disk atmospheres,  
if $\Htwo$ formation on grains is an important requirement for the 
formation of $\HtwoO$ 
in the warm disk atmosphere, 
the absence of grains 
in the inner disk region of a transition object may limit the 
$\HtwoO$ emission from this region.

Thus, the lack of strong MIR molecular emission from TW Hya indicates 
that the inner planet formation region of its gaseous disk 
($\lesssim 5$\,AU) 
has evolved away from the classical T Tauri norm. 
But further work is needed to determine whether that 
evolution is the result of the formation of a giant planet or   
other physical or chemical processes.

\subsection{Origin of the Detected Emission}
\subsubsection{Molecular Emission}

While strong molecular emission is not detected from TW Hya, we do 
detect a rich spectrum of weaker emission lines. 
Unlike the {\it Spitzer} spectrum of AA Tau (Carr \& Najita 2008), 
which shows a rich 
spectrum of $\HtwoO$, OH, HCN, $\CtwoHtwo$, and $\COtwo$ emission, 
and the strong $\HtwoO$ emission that is detected 
from the classical T Tauri stars DR Tau and AS 205A 
(Salyk et al.\ 2008), 
the molecular emission we detect from TW Hya in the 10--20$\micron$ 
region is dominated by radicals 
(OH, and possibly $\CHthree$) and ions ($\HCOp$).

Is the difference in the species detected in TW Hya 
vs.\ classical T Tauri stars 
due to the physical truncation of the disk, 
as discussed in the previous section?
Alternatively, or in addition, does the difference result from 
the higher X-ray and UV irradiation field of TW Hya? 
TW Hya has a nearly unique X-ray spectrum (Drake 2005) that 
shows unusually bright soft X-ray emission compared to other 
T Tauri stars (Robrade \& Schmitt 2006). 
TW Hya is also known as a bright FUV emission source (Herczeg et al.\ 2004). 
Bergin et al.\ (2004) quote an FUV flux (including Ly $\alpha$) of 
$G_0 = 3400$ at 100\,AU for TW Hya, in comparison with values of 
240, 340, and 1500 for DM Tau, GM Aur, and LkCa15, respectively. 
One hypothesis is that the stronger UV and soft X-ray field of TW Hya 
results from its lower accretion rate ($10^{-9}\Msunperyr$; 
Muzerolle et al.\ 2000; Alencar \& Basri 2000; Herczeg et al.\ 2004), 
which unburies 
the accretion shock and allows more energetic photons to emerge 
from the ``sides'' of the accretion column (Drake 2005; 
Ardila 2007; Ardila \& Johns-Krull 2009).

Perhaps this higher intensity photon field is responsible 
for 
some aspects of the emission that we see. 
A stronger UV field from TW Hya may dissociate $\HtwoO$ in favor of OH.
The hot OH emission that we detect may provide evidence 
for this process in action. 
This situation may therefore be similar to that of 
the Herbig Ae stars studied by  Mandell et al.\ (2008), 
who suggested that the presence of OH emission and lack of 
$\HtwoO$ emission from these sources could be the result of  
photodissociation by stellar UV. 

If it arises from the photodissociation of water, 
the hot OH emission we observe may help to explain the origin of the 
high OH column density that is inferred for the disk atmospheres of 
classical T Tauri stars. 
Carr \& Najita (2008) inferred a (line-of-sight) OH column density of 
$8 \times 10^{16}\persqcm$ at a temperature of $\sim 500$\,K 
based on their analysis of the {\it Spitzer} spectrum of AA Tau. 
For a system inclination of $i=75$ for AA Tau (Bouvier et al.\ 2007), 
the measured OH column density corresponds to a 
vertical column density of $2\times 10^{16}\,\persqcm$.
Even larger OH column densities of 
$2\times 10^{17}\persqcm$ 
were inferred for two classical T Tauri
stars based on $3\micron$ spectra (Salyk et al.\ 2008).

In comparison, Glassgold et al.\ (2009) find that X-ray 
irradiated disks will produce warm (300--1000\,K) OH 
(vertical) column densities of $1\times 10^{14} -8 \times 10^{14}\persqcm,$ 
with the specific value depending on the details of the model. 
Other chemical calculations for disks also fail to produce 
the large OH column densities (Agundez et al.\ 2008) 
or high OH/$\HtwoO$ abundances (Woods \& Willacy 2009) that are observed. 
Photodissociation of water to produce OH is one possible 
explanation for the much larger column density of OH 
that is observed 
(Bethell \& Bergin 2009; 
Glassgold et al.\ 2009). 
The hot OH emission that we observe may provide observational 
support for that perspective.

When water in the disk absorbs dissociating UV photons, the 
UV photons can both produce hot (prompt) OH emission as well as 
heat the disk atmosphere.  After it has emitted the prompt  
emission, the OH produced in this process may relax to a thermal 
distribution (at the gas temperature) and continue to emit thermally. 
A UV irradiated disk may thereby produce both thermal and 
prompt emission, as seen in a source such as AA Tau (section 3.5).  
Detailed modeling is needed to determine if the OH emission observed 
in a system like AA Tau is consistent in detail with this picture. 

Where would such water be located in the TW Hya system, 
given that no water 
emission is observed at 10--20$\micron$?  One possibility is that 
it resides in the optically thin region of the disk, but it 
is not excited enough or warm enough to emit 
in the 10--20$\micron$ region.  
Alternatively, the water 
may be present in the outer, optically thick disk 
where it is too cool to produce much emission at
10--20\,$\micron$. 
To address this issue, it would be interesting to measure 
velocity resolved OH line profiles 
in order to determine if the OH emission arises in a disk 
and at what radii.

The irradiation environment may also play a role in the 
$\HCOp$ emission that we detect. 
In their earlier study of organic molecules in the outer disk of 
TW Hya, Thi et al.\ (2004) found the 
abundance 
ratio of $\HCOp$/CO to be 
$\sim 10$ times larger 
in TW Hya than in two other transition disks (LkCa 15 and DM Tau).   
Thi et al.\ suggested that the high ratio of $\HCOp$/CO may result, 
in part, from the high X-ray flux of TW Hya. 
The high X-ray flux of TW Hya may also contribute to the prominence of 
$\HCOp$ emission in the mid-infrared.

\subsubsection{Atomic Emission}

As discussed in section 3.4, the [NeII] and [NeIII] fluxes 
we measure agree well with predictions for 
gaseous disks irradiated by stellar X-rays (see section 3.4).  
EUV irradiation of the disk may also contribute to some extent, 
although a soft UV spectrum is needed to be consistent with 
the large [NeII]/[NeIII] ratio observed. 
This result is consistent with the picture of a 
disk origin for the (spatially compact) [NeII] emission 
from T Tauri stars that is probed by high resolution spectroscopy. 
As discussed in Najita et al.\ (2009) 
based on high resolution spectroscopy of [NeII] emission 
from AA Tau and GM Aur, and Herczeg et al.\ (2007) in the context 
of TW Hya specifically, 
the [NeII] emission line is symmetric, centered at the radial 
velocity of the star, and has line widths that are plausibly 
explained by disk rotation.
The possibility of an origin in jets or outflows, illustrated in the 
case of the T Tau triplet (van Boekel et al.\ 2009), is less 
likely here, because TW Hya is not a known outflow source.  

This perspective is basically confirmed, although with a 
twist, in the more recent study of [NeII] in TW Hya using VISIR 
at the VLT.  Pascucci \& Sterzik (2009) 
interpret their measurement of the [NeII] line profile 
of TW Hya, which shows a $6\kms$ blueshift in the emission 
centroid, as indicating an origin in a low-velocity 
photoevaporative flow from the disk. 
This result is related to the idea of a disk origin for 
[NeII] emission in that the emission still arises from 
disk gas, although it is gas that is gravitationally 
unbound rather than in hydrostatic equilibrium. 

The HI emission we detect is not as well known or understood. 
While the $12.37\micron$ HI 7--6 line has been reported 
previously from T Tauri stars (e.g., Pascucci et al.\ 2007; 
see also Ratzka et al.\ 2007 for TW Hya), 
the weaker emission lines have not been reported previously. 
Perhaps such emission lines are commonly present in T Tauri stars, 
but they are obscured by stronger molecular emission and 
the weakness of the molecular emission in TW Hya makes the 
weaker HI lines easier to detect in this source.
On the other hand, TW Hya is an energetic UV and X-ray emission 
source, as discussed above, and the emission we detect may be 
more unusual as a result.   

What component of the T Tauri system could potentially 
produce the HI emission that we observe? 
Ratzka et al.\ (2007) suggested that 
the HI lines most likely originate from either an accretion shock 
or the stellar corona, although they were not specific 
on the details.  
Pascucci et al.\ (2007) have argued that the HI 6--5 and HI 7--6 
emission detected from other low accretion rate T Tauri stars are 
are too strong to be explained by 
magnetospheric accretion flows.  

Following the lead of previous studies that 
interpreted the Br$\gamma$ and Balmer lines of 
T Tauri stars as arising in a magnetospheric accretion flow 
(Calvet \& Hartmann 1992; Hartmann, Hewett, \& Calvet 1994;  
Najita et al.\ 1996; Muzerolle et al.\ 1998a,b; 
Folha \& Emerson 2001), 
Bary et al.\ (2008) interpreted the line ratios of the 
high-lying Paschen and Brackett lines they observed at low 
spectral resolution as arising in a 
magnetosphere with unusually low temperature.  
As described by Bary et al., 
the electron densities predicted theoretically for funnel 
flows (Martin 1996; Muzerolle et al.\ 2001) 
are consistent with their line ratios. 
They did not discuss whether magnetospheres can reproduce 
the fluxes of the Paschen and Brackett lines they 
observe.

The MIR HI lines impose further constraints on the origin of the 
HI emission.
It is unclear whether the HI emission we observe can be produced 
by the accretion shocks that 
are discussed in the literature. 
The accretion shock in the TW Hya system is expected to have 
a height 
of $\sim$ 1--1000\,km for the post-shock region 
(Ardila \& Johns-Krull 2009; Ardila 2007) 
and a density of $\sim 10^{13}\percc$  (Drake 2005; Ardila 2007). 
The pre-shock region in typical T Tauri stars is characterized 
as having a temperature of $\sim$ 10,000\,K, a height of $\sim 1000$ km, 
an average density of $10^{11}-10^{15}\percc$,  
and a filling factor of $f \sim 10^{-3} - 10^{-2}$ 
(Calvet \& Gullbring 1998).
These densities are high enough to produce inverted HI populations 
(Hummer \& Storey 1987).  
However, significant laser amplification may not occur or may 
be difficult to detect because of 
laser saturation and/or beaming effects (Strelnitzki et al.\ 1996; 
Smith et al.\ 1997).
Further work is needed to determine whether such high density, 
compact regions can produce bright HI emission with 
roughly case B line ratios. 

The line profiles of transitions connecting high lying HI levels 
may also provide some insights.  
One example is the Pa$\gamma$ (6--3) profile of TW Hya 
(Edwards et al.\ 2006), which is superficially similar to 
a magnetospheric accretion profile: 
the profile is steeper on the red side than the blue 
and the blue wing extends to $\sim 300\kms$. 
However, as noted by Edwards et al., the profile is not consistent 
in detail with predictions for magnetospheric accretion, 
e.g., the profile is narrower and more centered than predicted 
profiles (e.g., Muzerolle et al.\ 2001).  
Such narrow Pa$\gamma$ profiles are found commonly for 
low accretion rate systems (Edwards et al.\ 2006). 

Edwards et al.\ note a similarity with the narrow component 
of metallic emission lines (e.g., FeI and FeII), which are also 
the dominant component 
of the profile in low accretion sources such as TW Hya. 
The narrow and broad components are interpreted as arising 
in different physical regions, with the broad component  
having a significant contribution from funnel flows and 
an additional physically extended component from winds.  
The narrow component is associated 
with an accretion shock at the footpoint of the funnel flow 
(Edwards et al.\ 2006), which may not be the region 
from which the MIR HI emission originates, as noted above.

Disk atmospheres and/or photoevaporative flows may also contribute 
detectable HI emission.  
The disk atmosphere and photoevaporative flow that results from 
the X-ray irradiation of the disk, 
as described by Ercolano et al.\ (2008), 
produces a line luminosity of 
$4\times 10^{25}\,\ergpers$ in the HI 7--6 ($12.37\micron$) line. 
In an updated calculation, Ercolano et al.\ (2009) solve for the 
hydrostatic structure of the disk and include irradiation 
by both EUV and X-rays.  In this case, the HI 7--6 flux is revised 
upwards to $4\times 10^{26}\,\ergpers$ in FS0H0Lx1 
(B.\ Ercolano, private communication 2009), a model 
that is fairly optimistic in assuming no attenuation of the 
EUV flux by foreground HI absorption (e.g., as might be present 
in a magnetosphere). 
We observe a yet larger luminosity of $1.8\times 10^{28}\,\ergpers,$ 
the average of the flux in SH1 and SH2.  

%

A more general question is whether 
the neon and HI emission can 
arise from the same physical region of the TW Hya system.
The picture of [NeII] and [NeIII] emission arising in a mostly neutral 
region (e.g., in an X-ray irradiated disk atmosphere; section 3.4) 
contrasts, at least superficially, with the 
nominal picture of HI recombination line emission.  Case B 
hydrogen line ratios are typically interpreted as indicating formation 
in a fully ionized region.  
(While a high ionization fraction is clearly not required to produce 
recombination emission, it helps to maximize the emission from a 
given mass of gas.)

The mismatch in these conditions seems to show up in other models of 
EUV and X-ray irradiated disks as well. 
Hollenbach \& Gorti (2009) 
show that although 
irradiation by soft EUV 
can account for the flux of the [NeII] emission observed from 
T Tauri stars, 
it cannot produce the observed 
HI 7--6 line luminosity: the 
predicted 
HI 7--6 flux is $<1$\% of the [NeII] flux. 
As a result, 
Hollenbach \& Gorti (2009) argue that the HI emission arises 
from a different region than the [NeII]. 
As they note, a high density region 
is one possibility since it may produce significant HI 
emission without producing additional (unwanted) [NeII].  
Considerations of HI population inversions and laser amplification, 
and whether the HI emission would have (optically thin) 
case B line ratios, as described above, may limit the 
allowable range of densities. 

Given the current uncertainty regarding the possible origin of the HI 
emission, it appears that 
high resolution spectroscopy of the MIR HI lines would likely provide 
valuable, and much needed, insights into the origin of the emission. 
Since the $11.31\micron$ and $12.37\micron$ lines are comparable in 
brightness to the [NeII] line, such a study should be feasible 
currently. 
If the mid-infrared lines are found to be as broad as the Pa$\gamma$
line, the role of high velocity phenomena such as magnetospheric 
infall would be indicated.  
If the lines are narrow ($>30 \kms$) and/or with a small blueshift
($\sim 10\kms$), disk atmospheres and/or photoevaporative flows would 
be indicated.  

Both components might in fact contribute.  
As noted above, the Pa$\gamma$ profiles of Edwards et al.\ (2006) are 
narrower and more centered in velocity than magnetospheric emission models.  
A combination of a high-velocity magnetospheric component and a 
low velocity disk/photoevaporative component may better account 
for the observed profiles than a magnetosphere alone.

\section{Summary and Future Directions}

The {\it Spitzer} spectrum of TW Hya is strikingly different from that 
of other classical T Tauri stars reported in the literature, 
showing weak molecular emission in the $10-20\micron$ region 
and displaying no strong emission features of $\HtwoO$, 
$\CtwoHtwo,$ or HCN.  
The difference indicates that the inner planet formation region 
($\lesssim 5$\,AU) of its gaseous disk has evolved away 
from the classical T Tauri norm. 
With the SED of TW Hya indicating that the dust disk within $\sim 4$\,AU 
of the star has become optically thin in the continuum, possibly 
because it has been cleared by an orbiting giant planet (Calvet et al.\ 2002), 
it is tempting to infer that TW Hya shows weak molecular emission 
in the mid-infrared 
because the planet has created a gap in the {\it gaseous} disk as well. 
We discussed in section 4 some of the issues that need to be investigated 
in order to determine whether the difference we observe is the result of 
giant planet formation or other processes. 

Given our limited understanding of the factors that govern 
the emission spectra of T Tauri disks, it may 
be useful to take an empirical approach in exploring whether 
the lack of molecular emission seen from TW Hya is a consequence 
of the physical evolution of the disk, its chemical evolution,  
or an excitation effect.  
For example, to explore the possibility of an excitation effect, 
we can compare  the {\it Spitzer} spectrum of TW Hya with those of 
other non-transition objects with comparable accretion rates. 
If non-transition objects with low accretion rates also lack 
strong molecular emission, the spectrum of TW Hya would not be 
unusual for its accretion rate and the lack of molecular emission 
may be the result of poor excitation. 
However, if strong molecular emission is observed in other low 
accretion rate 
systems, the absence of such emission in TW Hya would suggest 
a gap in its gaseous disk or a possible chemical effect. 
We will take this approach in a future study. 

Similarly, it would be interesting to explore whether other 
transition objects also show weak molecular emission compared to 
classical T Tauri stars.   If transition objects with much weaker UV fluxes 
than that of TW Hya (e.g., DM Tau, GM Aur) also show a deficit of 
molecular emission, that would suggest a more dominant role for 
SED evolution (a deficit of grains; possible clearing by a giant planet), 
rather than photochemistry, in accounting for the difference in the spectra.
We will report on this in future publications.  

While strong molecular emission is not detected from TW Hya, we do 
detect a rich spectrum of emission lines of atoms (HI, [NeII], and [NeIII]) and 
molecules ($\Htwo,$ OH, $\COtwo,$ $\HCOp,$ and possibly $\CHthree$). 
One of the most intriguing is the OH emission, which is hot and 
may result from the UV photodissociation of water. 
A more detailed analysis of the OH emission spectrum may be 
able to determine whether it is produced by photodissociation. 
The properties of the molecular emission from TW Hya, both the OH 
and other molecules, will be analyzed in greater detail in a future study.  


Because we detect multiple HI lines, we can show that the HI emission 
from TW Hya has a recombination spectrum. 
In contrast to the neon emission from TW Hya, which can be 
well accounted for by (primarily) stellar X-ray irradiation of the disk, 
the physical origin of the HI emission is difficult to identify.  As discussed 
in section 4, magnetospheres, disk atmospheres and/or photoevaporative 
flows could plausibly contribute to the emission and multiple components 
may play a role.  High resolution spectroscopy 
of the brightest HI lines (HI 7-6 and HI 9-7) would likely provide 
valuable insights into the origin of the emission.






\acknowledgments

We are grateful to Al Glassgold, Barbara Ercolano, and David Ardila for 
interesting and useful discussions regarding the interpretation 
of the observations. 
JN thanks for their generous hospitality Tom Soifer and the 
{\it Spitzer} Science Center, where much of the analysis for 
this paper was carried out. 
This work is based on observations made with the {\it Spitzer Space
Telescope}, which is operated by the Jet Propulsion Laboratory,
California Institute of Technology under a contract with NASA.
Support for this work was provided by NASA through an award issued
by JPL/Caltech.  Basic research in infrared astronomy at the Naval
Research Laboratory is supported by 6.1 base funding.



{\it Facilities:} \facility{Spitzer Space Telescope (IRS)}

\bigskip
  
\begin{figure}
\figurenum{1}
\plotfiddle{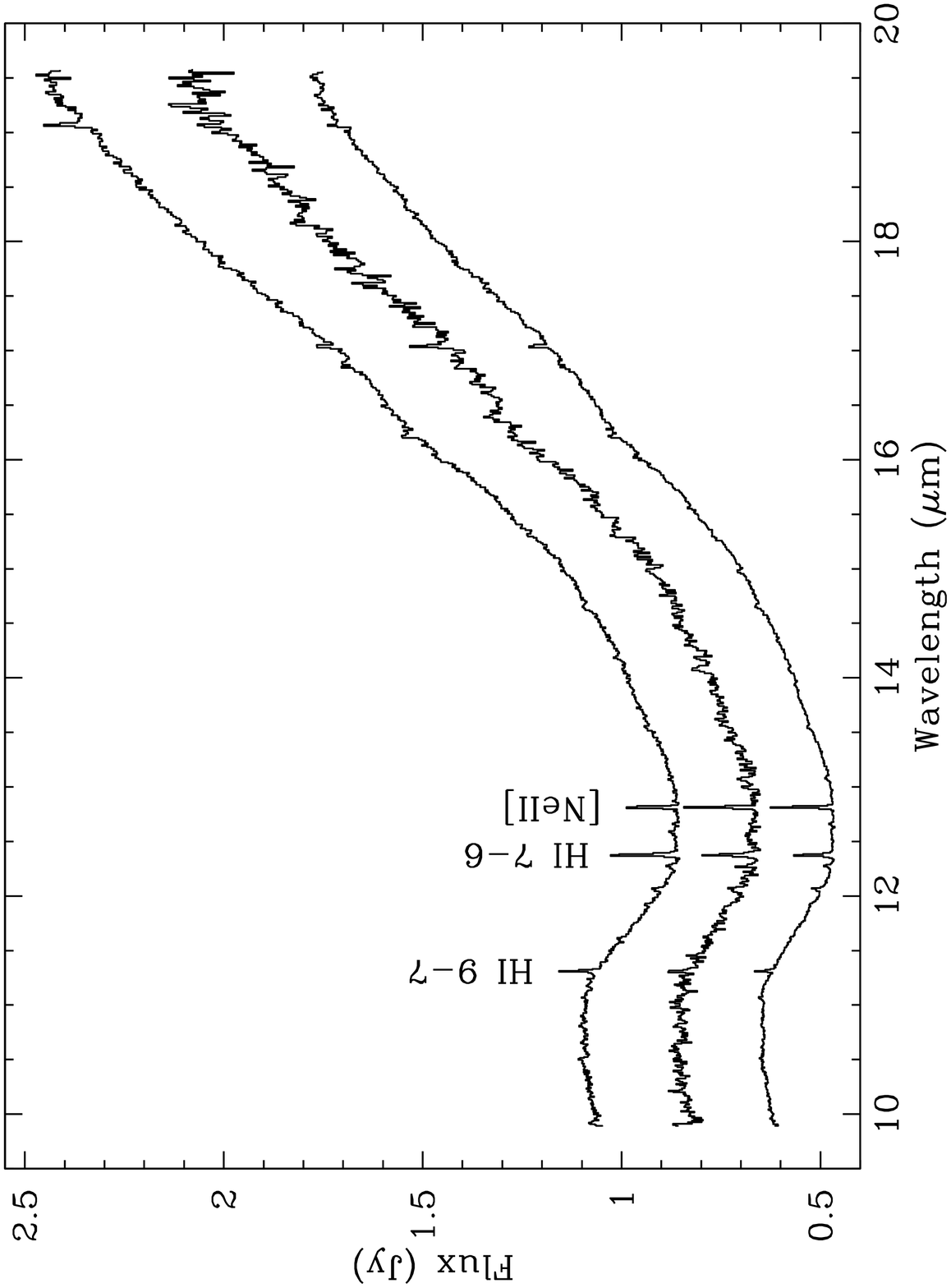}{5.5truein}{270}{60}{60}{-235}{400}
\caption{{\it Spitzer} IRS short-high spectra of TW Hya taken at three epochs: SH1 (top), SH2 (bottom), 
and the GTO spectrum (middle).  The GTO spectrum is offset by $+0.15$ Jy and 
SH1 by $+0.25$ Jy for clarity. 
}
\end{figure}

\begin{figure}
\figurenum{2}
\plotfiddle{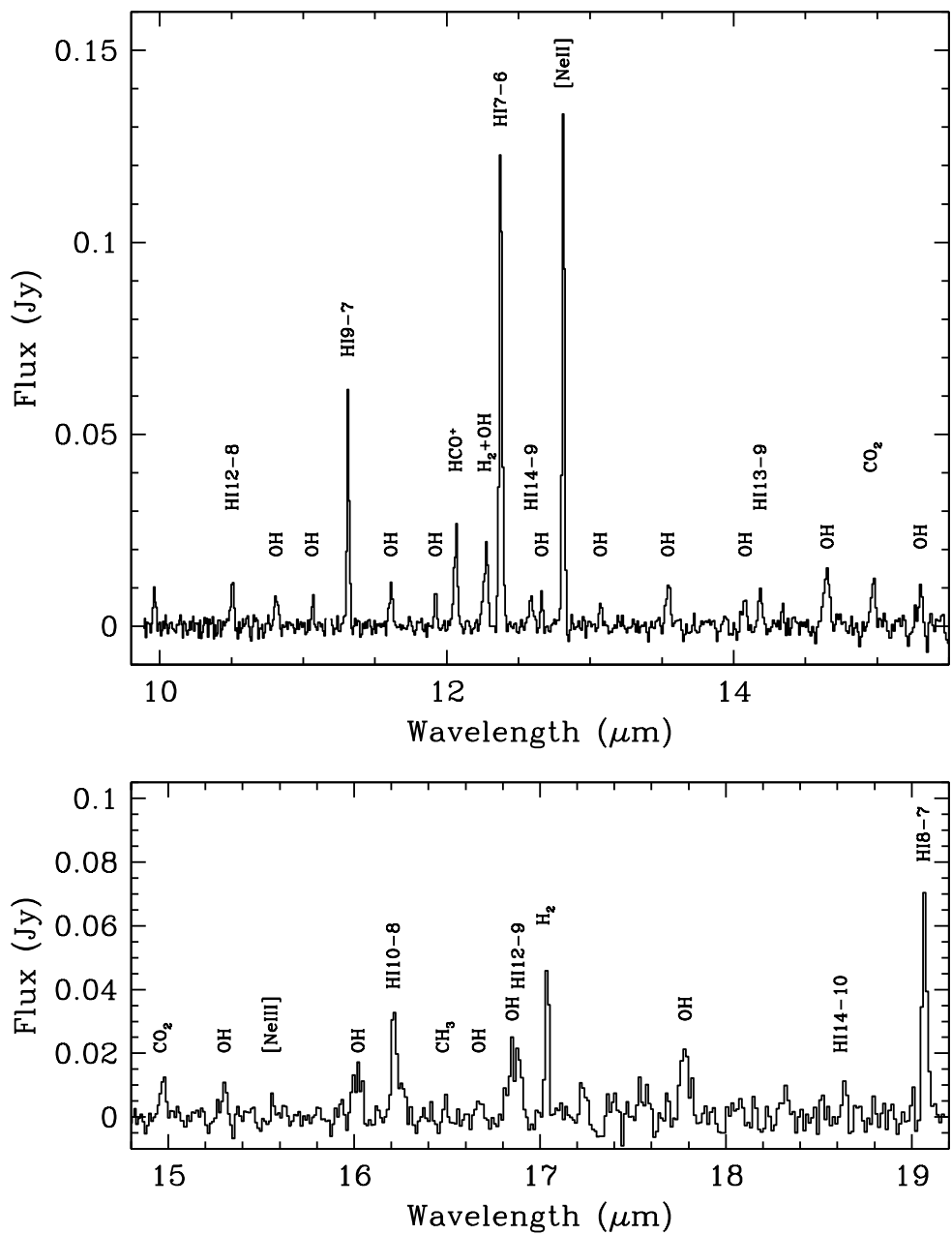}{6.0truein}{0}{70}{70}{-210}{-25}
\caption{Average of the SH1 and SH2 spectra of TW Hya 
from which a polynomial fit to the continuum has been subtracted.  
Emission features of atomic lines (HI, [NeII], and possibly [NeIII]) and 
molecules ($\Htwo$, OH, $\COtwo$, $\HCOp$, and possibly $\CHthree$) 
are detected.  The [NeIII] and $\CHthree$ features are clearer
in the SH2 and SH1 spectra, respectively (see Figure 3).
}
\end{figure}

\begin{figure}
\figurenum{3}
\plotfiddle{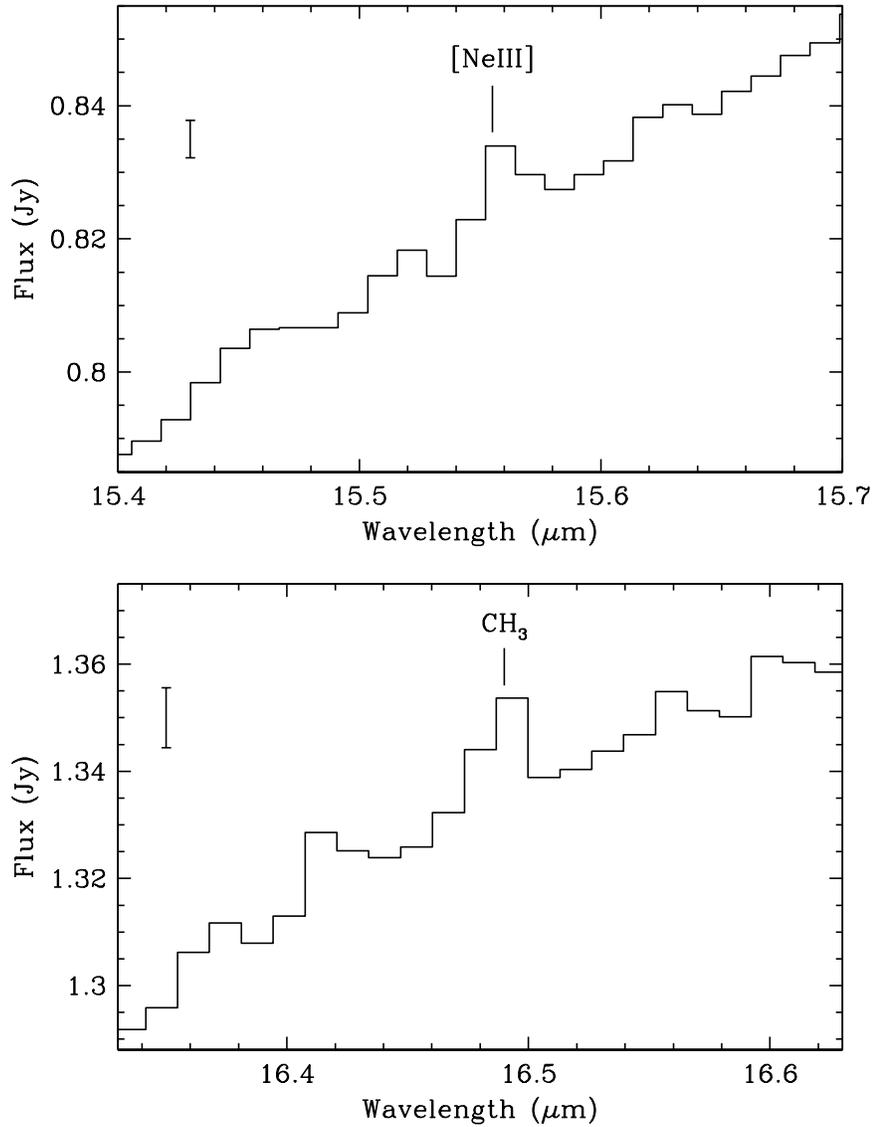}{6.0truein}{0}{60}{60}{-190}{25}
\caption{The region around the [NeIII] line in the SH2 spectrum 
(top) and the $\CHthree$ emission band in the SH1 spectrum (bottom).
A representative errorbar ($\pm$ 1-$\sigma$) is shown in the upper left 
corner of each panel. 
}
\end{figure}


\begin{figure}
\figurenum{4}
\plotfiddle{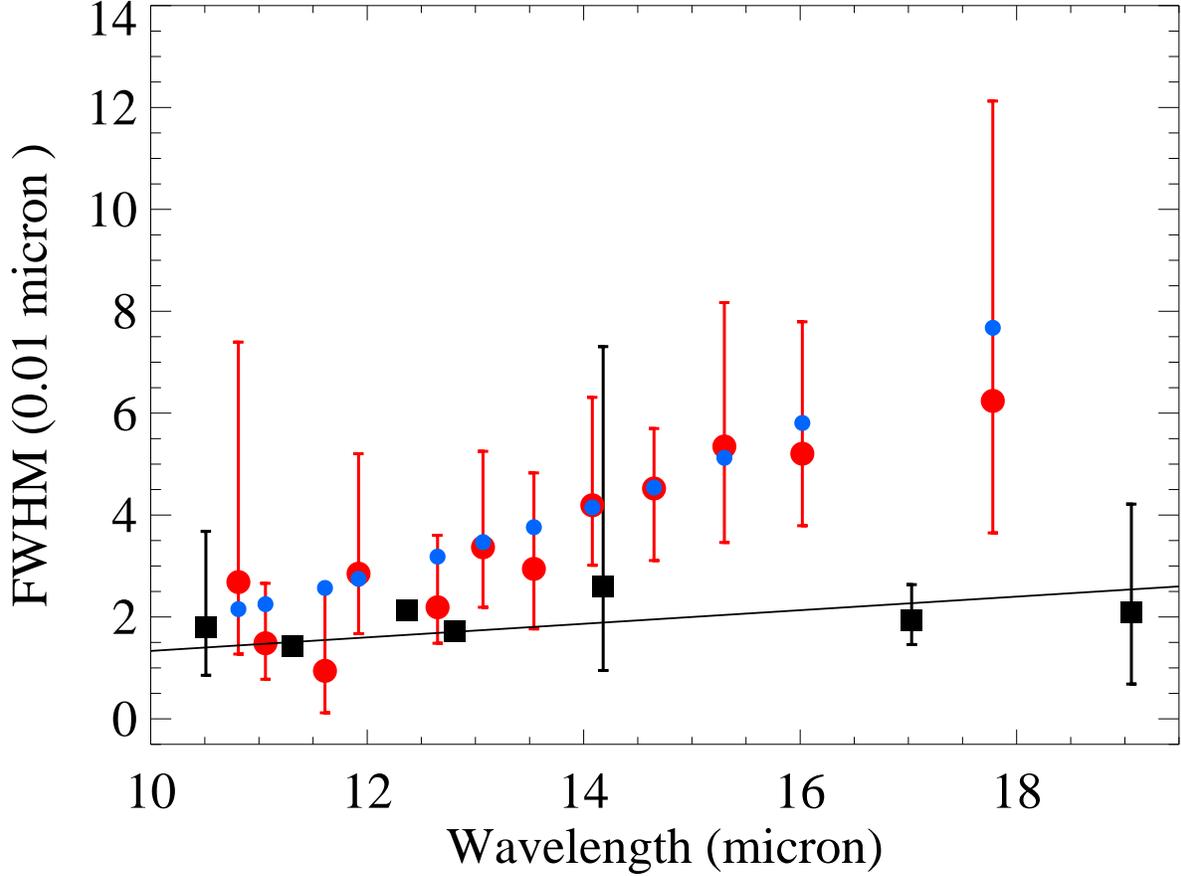}{5.5truein}{0}{80}{80}{-270}{50}
\caption{FWHM of the Gaussian fits to OH features (large red dots) and 
[NeII] and unblended HI lines (black squares) in the SH2 spectrum.
Where errorbars are not apparent, they are smaller than the size of the symbol. 
The nominal resolution of IRS in the short-high mode ($R=750$; black line) is shown 
for comparison. 
While the HI and [NeII] lines are spectrally unresolved, 
the widths of the OH features grow with wavelength due to the increasing 
spread in the wavelengths of individual OH lines that make up a given feature. 
This is illustrated by the agreement with the small blue dots, which show this 
spread in wavelengths added in quadrature with the spectral resolution at the 
central wavelength of each feature. 
}
\end{figure}

\begin{figure}
\figurenum{5}
\plotfiddle{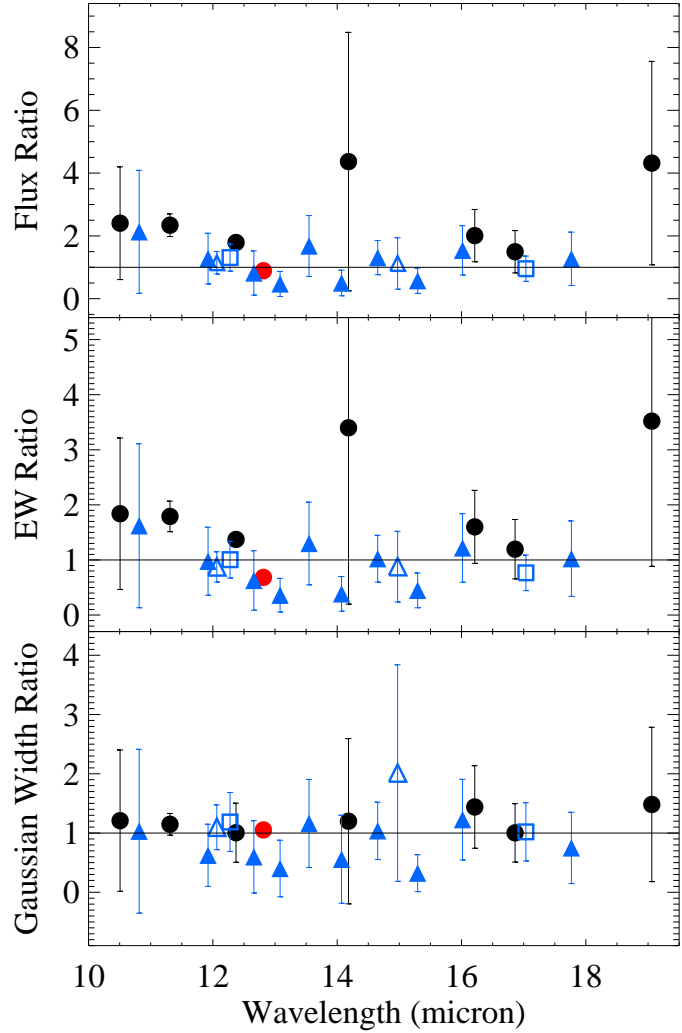}{6.5truein}{0}{80}{80}{-150}{50}
\caption{Ratio of line fluxes (top), equivalent widths (middle), and Gaussian FWHM (bottom)
in SH1 compared to SH2 for HI (black dots), [NeII] (red dot), 
$\Htwo$ (open blue squares), OH (solid blue triangles), and $\HCOp$ and $\CHthree$ 
(open blue triangles). 
The fluxes of the HI lines are larger by $\sim 2$ in flux in SH1 compared to SH2.
The equivalent width of the [NeII] line is smaller in SH1 than SH2. 
}
\end{figure}

\begin{figure}
\figurenum{6}
\plotfiddle{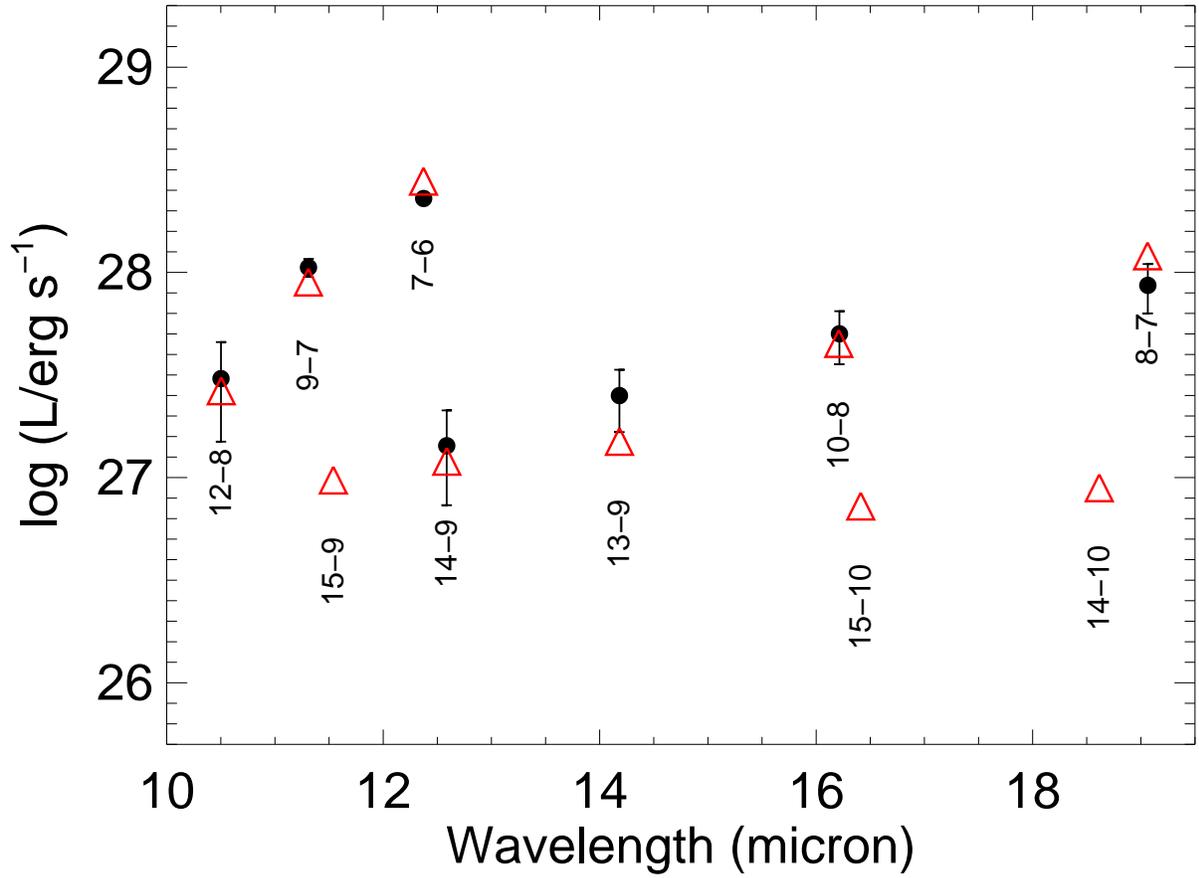}{5.5truein}{0}{80}{80}{-270}{0} 
\caption{HI line luminosities from the SH1 spectrum (dots with error bars) compared with 
relative fluxes from case B recombination (triangles).  
The HI 15-9, HI 15-10, and HI 14-10 
lines are not detected, consistent with their low anticipated fluxes. 
}
\end{figure}

\begin{figure}
\figurenum{7}
\plotfiddle{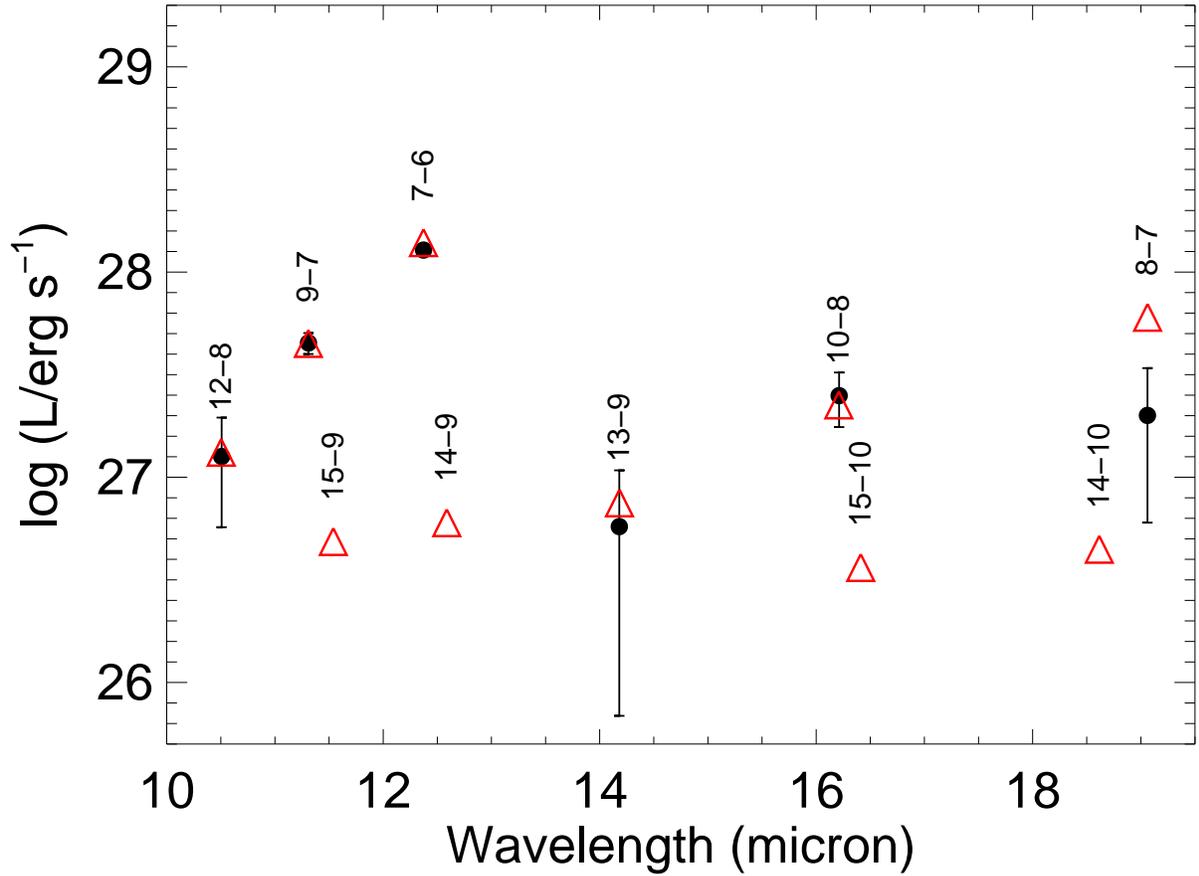}{5.5truein}{0}{80}{80}{-270}{0} 
\caption{HI line luminosities from the SH2 spectrum (dots with error bars) compared with 
relative fluxes from case B recombination (triangles). 
The HI 8-7 line is weaker at this epoch and not in as good agreement with 
the case B relative fluxes.
The HI 15-9, HI 14-9, HI 15-10, and  HI 14-10 
lines are not detected, consistent with their low anticipated fluxes. 
}
\end{figure}

\begin{deluxetable}{llrrrr} 
\tablecolumns{6} 
\tablewidth{0pc} 
\tablecaption{Observing Log for TW Hya SH Datasets} 
\tablehead{ 
\colhead{Dataset}  & 
\colhead{Date}  & 
\colhead{AOR}  & 
\colhead{Ramp}  & 
\colhead{On-Source}  & 
\colhead{Off-Source}  \\ 
\colhead{}  & 
\colhead{}  & 
\colhead{}  & 
\colhead{(sec)}  & 
\colhead{(sec)}  & 
\colhead{(sec)}   
}
\startdata 
GTO &   2004-01-04 & 3571456    &  6  &      12 &        0 \\
SH1 &   2006-07-01 & 18017792\tablenotemark{*}  & 30  &     120 &      120 \\
SH2 &   2008-01-07 & 24402944   & 30  &     600 &      300 \\
\enddata 
\tablenotetext{*}{AOR for sky observations: 18018048, 18018304.}
\end{deluxetable}

\begin{deluxetable}{lccccccc} 
\tablecolumns{8} 
\tablewidth{0pc} 
\tablecaption{Continuum and Errors} 
\tablehead{ 
\colhead{}  & 
\colhead{}  & 
\multicolumn{2}{c}{GTO}& 
\multicolumn{2}{c}{SH1}& 
\multicolumn{2}{c}{SH2}  \\
\cline {3-4} \cline {5-6} \cline {7-8}  \\
\colhead{Feature}  & 
\colhead{Wavelength}  & 
\colhead{Continuum}  & 
\colhead{Error}  & 
\colhead{Continuum}  & 
\colhead{Error}  & 
\colhead{Continuum}  & 
\colhead{Error}  \\ 
\colhead{}  & 
\colhead{($\mu$m)}  & 
\colhead{(Jy)}  & 
\colhead{(Jy)}  & 
\colhead{(Jy)}  & 
\colhead{(Jy)}  & 
\colhead{(Jy)}  & 
\colhead{(Jy)}   
}
\startdata 
HI 9-7  & 11.31  &   0.68   &  0.013   &  0.81    &  0.0037   &  0.62  &   0.0019   \\
NeII    & 12.81  &   0.51   &  0.011   &  0.61    &  0.0030   &  0.47  &   0.0017   \\
$\COtwo$& 14.97  &   0.76   &  0.015   &  0.89    &  0.0036   &  0.69  &   0.0019   \\
H2 S(1) & 17.03  &   1.28   &  0.023   &  1.47    &  0.0063   &  1.18  &   0.0050   \\
HI 8-7  & 19.06  &   1.85   &  0.029   &  2.09    &  0.012    &  1.70  &   0.0084   \\
\enddata 
\end{deluxetable}

\begin{deluxetable}{ccc} 
\tablecolumns{3} 
\tablewidth{0pc} 
\tablecaption{OH Features Detected} 
\tablehead{ 
\colhead{$\lambda/\mu$m}  & 
\colhead{$v$}  & 
\colhead{Transitions} 
}
\startdata 
10.81	&	0 &	3/2 R 29.5, 1/2 R 28.5 \\
11.06	&	0 &	3/2 R 28.5, 1/2 R 27.5 \\
11.31	&	0 &	3/2 R 27.5, 1/2 R 26.5 \\
11.61	&	0 &	3/2 R 26.5, 1/2 R 25.5 \\
11.92	&	0 &	3/2 R 25.5, 1/2 R 24.5 \\
12.65	&	0 &	3/2 R 23.5, 1/2 R 22.5 \\
13.08	&	0 &	3/2 R 22.5, 1/2 R 21.5 \\
13.54	&	0 &	3/2 R 21.5, 1/2 R 20.5 \\
14.08	&	0 &	3/2 R 20.5, 1/2 R 19.5 \\
14.65	&	0 &	3/2 R 19.5, 1/2 R 18.5 \\
	&	1 &	3/2 R 20.5, 1/2 R 19.5 \\
15.30	&	0 &	3/2 R 18.5, 1/2 R 17.5 \\
	&	1 &	3/2 R 19.5, 1/2 R 18.5 \\
16.02	&	0 &	3/2 R 17.5, 1/2 R 16.5 \\
16.84\tablenotemark{*}	&	0 &	3/2 R 16.5, 1/2 R 15.5 \\ 
17.78	&	0 &	3/2 R 15.5, 1/2 R 14.5 \\
\enddata 
\tablenotetext{*}{Blended with the HI 12-9 line.}
\end{deluxetable}

\begin{deluxetable}{lccccccccc} 
\tablecolumns{9} 
\tablewidth{0pc} 
\tablecaption{Emission Line Properties in the SH1 Spectrum} 
\tablehead{ 
\colhead{Feature}  & 
\colhead{$\lambda$}   & 
\colhead{Flux} & 
\colhead{err} & 
\colhead{FWHM}& 
\colhead{$+$err}  & 
\colhead{$-$err}  & 
\colhead{EW} &
\colhead{err} \\ 
\colhead{}  & 
\colhead{($\micron$)}   & 
\multicolumn{2}{l}{($10^{-14}\ergperssqcm$)} & 
\multicolumn{3}{l}{($10^{-2}\micron$)}& 
\multicolumn{2}{l}{($10^{-3}\micron$)} 
}
\startdata 
%
HI 12-8  & 10.502  & 1.07   & 0.54   & 2.2   & 2.1   & 0.9   & 0.47  & 0.24  \\
OH       & 10.809  & 0.82   & 0.51   & 2.8   & 3.5   & 1.9   & 0.38  & 0.24  \\
OH       & 11.077  & 1.26   & 0.73   & 5.1   & 5.2   & 2.4   & 0.62  & 0.36  \\
HI 9-7   & 11.310  & 3.74   & 0.38   & 1.6   & 0.2   & 0.1   & 1.96  & 0.20  \\
OH       & 11.609  & 0.69   & 0.40   & 2.0   & 1.9   & 1.2   & 0.42  & 0.24  \\
OH       & 11.922  & 0.66   & 0.30   & 1.8   & 1.6   & 0.7   & 0.46  & 0.21  \\
$\HCOp$  & 12.064  & 1.51   & 0.41   & 2.8   & 1.1   & 0.6   & 1.13  & 0.31  \\
H$_2$S(2)+OH& 12.276&1.49   & 0.42   & 3.4   & 1.4   & 0.9   & 1.22  & 0.34  \\
HI 7-6   & 12.374  & 8.10   & 0.33   & 2.1   & 0.9   & 1.2   & 6.75  & 0.28  \\
HI 14-9  & 12.587  & 0.51   & 0.25   & 1.5   & 1.6   & 0.7   & 0.43  & 0.21  \\
OH       & 12.663  & 0.37   & 0.28   & 1.3   & 2.0   & 0.7   & 0.33  & 0.25  \\
NeII     & 12.814  & 4.95   & 0.29   & 1.8   & 0.1   & 0.1   & 4.43  & 0.26  \\
OH       & 13.069  & 0.26   & 0.19   & 1.3   & 2.4   & 0.9   & 0.23  & 0.18  \\
OH       & 13.541  & 0.80   & 0.33   & 3.4   & 1.9   & 0.9   & 0.73  & 0.30  \\
OH       & 14.074  & 0.31   & 0.23   & 2.3   & 4.7   & 1.9   & 0.28  & 0.20  \\
HI 13-9  & 14.183  & 0.89   & 0.30   & 3.1   & 1.6   & 1.2   & 0.79  & 0.27  \\
OH       & 14.645  & 1.23   & 0.42   & 4.7   & 2.1   & 1.4   & 1.06  & 0.36  \\
CO$_2$   & 14.974  & 0.70   & 0.47   & 5.4   & 6.1   & 3.5   & 0.59  & 0.39  \\
OH       & 15.302  & 0.43   & 0.26   & 1.7   & 2.4   & 0.9   & 0.34  & 0.21  \\
OH       & 16.013  & 1.56   & 0.59   & 6.4   & 4.2   & 1.6   & 1.12  & 0.42  \\
HI 10-8  & 16.215  & 1.78   & 0.51   & 3.8   & 1.4   & 1.2   & 1.23  & 0.36  \\
$\CHthree$& 16.482 & 0.45   & 0.33   & 1.9   & 4.7   & 1.6   & 0.30  & 0.23  \\
HI12-9+OH& 16.859  & 2.53   & 0.74   & 8.0   & 3.5   & 2.1   & 1.69  & 0.49  \\
H$_2$ S(1)& 17.036 & 1.21   & 0.41   & 2.0   & 1.2   & 0.5   & 0.79  & 0.27  \\
OH       & 17.762  & 1.38   & 0.57   & 4.7   & 2.6   & 2.1   & 0.84  & 0.35  \\
HI 8-7   & 19.064  & 3.06   & 0.83   & 3.1   & 0.9   & 0.9   & 1.77  & 0.48  \\
\enddata 
\end{deluxetable} 

\begin{deluxetable}{lccccccccc} 
\tablecolumns{9} 
\tablewidth{0pc} 
\tablecaption{Emission Line Properties in the SH2 Spectrum} 
\tablehead{ 
\colhead{Feature}  & 
\colhead{$\lambda$}   & 
\colhead{Flux} & 
\colhead{err} & 
\colhead{FWHM} & 
\colhead{$+$err}  & 
\colhead{$-$err}  & 
\colhead{EW} &
\colhead{err} \\
\colhead{}  & 
\colhead{($\micron$)}   & 
\multicolumn{2}{l}{($10^{-14}\ergperssqcm$)} & 
\multicolumn{3}{l}{($10^{-2}\micron$)}& 
\multicolumn{2}{l}{($10^{-3}\micron$)} 
}
\startdata 
%
%
HI 12-8  & 10.508  & 0.45   & 0.24   & 1.8   & 1.9   & 0.9   & 0.26  & 0.14  \\
OH       & 10.8146 & 0.38   & 0.26   & 2.7   & 4.7   & 1.4   & 0.23  & 0.16  \\
OH       & 11.063  & 0.43   & 0.20   & 1.5   & 1.2   & 0.7   & 0.27  & 0.12  \\
HI 9-7   & 11.310  & 1.60   & 0.19   & 1.4   & 0.2   & 0.2   & 1.10  & 0.13  \\
OH       & 11.614  & 0.15   & 0.13   & 0.9   & 1.6   & 0.8   & 0.12  & 0.11  \\
OH       & 11.922  & 0.52   & 0.23   & 2.9   & 2.4   & 1.2   & 0.47  & 0.21  \\
$\HCOp$  & 12.064  & 1.32   & 0.21   & 2.6   & 0.5   & 0.6   & 1.30  & 0.20  \\
H$_2$S(2)+OH&12.276& 1.14   & 0.20   & 2.9   & 0.7   & 0.7   & 1.21  & 0.21  \\
HI 7-6   & 12.373  & 4.53   & 0.20   & 2.1   & 0.1   & 0.1   & 4.92  & 0.21  \\
OH       & 12.659  & 0.45   & 0.19   & 2.2   & 1.4   & 0.7   & 0.52  & 0.21  \\
NeII     & 12.813  & 5.56   & 0.16   & 1.72  & 0.0   & 0.1   & 6.47  & 0.19  \\
OH       & 13.081  & 0.55   & 0.21   & 3.4   & 1.9   & 1.2   & 0.65  & 0.25  \\
OH       & 13.547  & 0.48   & 0.20   & 2.9   & 1.9   & 1.2   & 0.56  & 0.23  \\
OH       & 14.070  & 0.62   & 0.23   & 4.2   & 2.1   & 1.2   & 0.72  & 0.26  \\
HI 13-9  & 14.181  & 0.20   & 0.18   & 2.6   & 4.7   & 1.6   & 0.23  & 0.21  \\
OH       & 14.651  & 0.94   & 0.22   & 4.5   & 1.2   & 1.4   & 1.04  & 0.25  \\
CO$_2$   & 14.972  & 0.62   & 0.17   & 2.7   & 1.1   & 0.6   & 0.67  & 0.19  \\
OH       & 15.294  & 0.75   & 0.28   & 5.4   & 2.8   & 1.9   & 0.77  & 0.29  \\
NeIII    & 15.558  & 0.25   & 0.19   & 1.6   & 2.1   & 1.2   & 0.24  & 0.18  \\
OH       & 16.016  & 1.01   & 0.35   & 5.2   & 2.6   & 1.4   & 0.92  & 0.32  \\
HI 10-8  & 16.212  & 0.88   & 0.26   & 2.6   & 1.2   & 0.7   & 0.77  & 0.23  \\
HI12-9+OH& 16.861  & 1.69   & 0.58   & 8.0   & 3.8   & 2.1   & 1.41  & 0.48  \\
H$_2$ S(1)& 17.035 & 1.26   & 0.31   & 1.9   & 0.7   & 0.5   & 1.03  & 0.25  \\
OH       & 17.767  & 1.08   & 0.57   & 6.2   & 5.9   & 2.6   & 0.82  & 0.43  \\
HI 8-7   & 19.061  & 0.71   & 0.50   & 2.1   & 2.1   & 1.4   & 0.50  & 0.35  \\
\enddata 
\end{deluxetable}

\begin{deluxetable}{llcccc} 
\tablecolumns{6} 
\tablewidth{0pc} 
\tablecaption{Emission Line Properties in the GTO Spectrum} 
\tablehead{ 
\colhead{Feature}  & 
\colhead{$\lambda$}   & 
\colhead{Flux} & 
\colhead{err} & 
\colhead{EW} &
\colhead{err}  \\
\colhead{}  & 
\colhead{($\micron$)}   & 
\multicolumn{2}{l}{($10^{-14}\ergperssqcm$)} & 
\multicolumn{2}{l}{($10^{-3}\micron$)} 
}
\startdata 
%
HI 9-7   & 11.306  & 2.92   & 1.53   & 1.83  & 0.96  \\
$\HCOp$  & 12.060  & 2.39   & 1.44   & 2.12  & 1.28  \\
H$_2$ S(2)& 12.272 & 1.55   & 1.00   & 1.51  & 0.97  \\
HI 7-6   & 12.372  & 7.13   & 1.37   & 7.13  & 1.37  \\
NeII     & 12.812  & 5.90   & 0.91   & 6.29  & 0.97  \\
\enddata 
\end{deluxetable} 

\end{document}